\documentclass{jfm}
\usepackage{graphicx}
\usepackage{epstopdf, epsfig}
\usepackage{amsmath}
\usepackage{bm}

\shorttitle{Experiments on jet in a crossflow in low-velocity-ratio regime}
\shortauthor{L. Klotz, K. Gumowski and J.E. Wesfreid}
\title{Experiments on jet in a crossflow in low-velocity-ratio regime}%

\author{L. Klotz\aff{1,2}
  \corresp{\email{lukasz.klotz@ist.ac.at}},
  K. Gumowski\aff{2} and J.E. Wesfreid\aff{3}
  }

\affiliation{ \aff{1} Institute of Science and Technology, Am Campus 1, 3400 Klosterneuburg, Austria \aff{2} Institute of Aeronautics and Applied Mechanics, Warsaw University of Technology, Nowowiejska 24, 00-665 Warsaw, Poland \aff{3}Physique et M\'ecanique des Milieux H\'et\'erog\`enes (PMMH), CNRS - ESPCI - PSL Research University, 10 rue Vauquelin, 75005 Paris, France; Sorbone Universit\'e; Universit\'e Paris Diderot}
\graphicspath{{./Fig/}}

\begin{document}

\maketitle

\begin{abstract}

The hairpin instability of a jet in a crossflow (JICF) for low jet-to-crossflow velocity ratio is investigated experimentally for a velocity ratio range of $R\in(0.14,0.75)$ and crossflow Reynolds numbers $\Rey_D\in(260,640)$. From spectral analysis, we characterize the Strouhal number and amplitude of the hairpin instability as a function of $R$ and $\Rey_D$. We demonstrate that the dynamics of the hairpins is well described by the Landau model, and hence that the instability occurs through Hopf bifurcation, similarly to other hydrodynamical oscillators such as wake behind different bluff bodies. Using the Landau model, we determine the precise threshold values of hairpin shedding. We also study the spatial dependence of this hydrodynamical instability, which shows a global behaviour.

\end{abstract}
\section{Introduction}
Jet in a crossflow (JICF) refers to the semi-bounded flow in which a flux of fluid from an orifice in a lower bounding wall interacts with the main flow above. Here we consider the canonical configuration of a round jet orifice oriented perpendicularly to the crossflow and flush with the lower bounding wall. This is commonly entountered in both natural and engineering systems, such as combustion chambers, chemical mixing, volcanic eruptions, pollutant discharges or film cooling. The dynamics of JICF is primarily dictated by the velocity ratio between the jet and free-stream. The velocity ratio can be defined with either bulk or centerline speed of the jet, to which we will refer as $R$ and $R^*$, respectively. 

JICF can be categorized in two big categories depending on the velocity ratio. The first category contains jets with high velocity ratio, which have been extensively investigated over several decades \citep[see reviews of][]{margason_fifty_1993, karagozian_transverse_2010, mahesh_interaction_2013,karagozian_jet_2014}. Their topology consists of shear-layer vortices, counter-rotating vortex pairs, upright wake vortices, and horseshoe vortices. \citet[][]{megerian_transverse_jet_2007} investigated flush and elevated jet nozzles and reported that for low enough $R$  self-sustained strong pure-tone oscillations are generated by rolling-up of the upper jet shear layer very close to the leading edge of the jet orifice. These observations were later confirmed by other experiments \citep{davitian_transition_2010,getsinger_structural_2014,gevorkyan_influence_2018}, as well as numerically \citep{iyer_numerical_2016} and theoretically \citep{regan_global_2017}. 

\begin{figure}

\hspace{1.65cm}\includegraphics[scale=0.75]{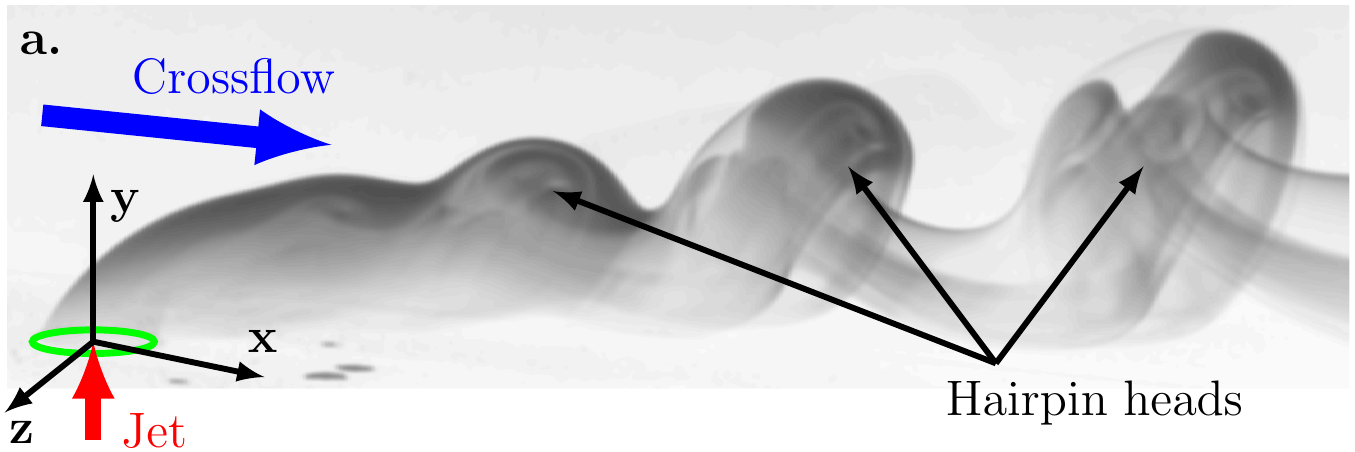}\\
\caption{ a) Flow visualisation (streaklines) to show schematically flow structure of hairpin vortices in JICF for the low-velocity-ratio regime. The jet orifice, crossflow, and jet flux are marked by green ellipse, blue arrow and red arrow, respectively. Additionally, hairpin vortices are indicated by black arrows. \vspace{0.4cm}}
\label{fig:Schematic}    
\end{figure}

On the other hand, JICF with low velocity of injection is less well described. Experiments \citep{gopalan_structure_2004} and numerical simulations \citep{sau_dynamics_2008} indicate in this case that the intensity of the upstream jet shear layer is significantly decreased due to the interaction with the incoming crossflow. In addition, numerous experiments \citep{acarlar_study_1987,lim_development_2001,camussi_experimental_2002,bidan_steady_2013,cambonie_transition_2014} reported about periodic hairpin shedding (Fig. \ref{fig:Schematic}), the heads of which consist of the spanwise vorticity (measured in the symmetry plane of JICF) of the same/opposite sign when compared to the downstream/upstream jet shear layer. \citet{camussi_experimental_2002} observed for a single crossflow Reynolds number (defined with free-stream velocity, jet diameter and kinematic viscosity of the fluid) that with increasing $R$, the amplitude of the vorticity emanating from the upstream jet shear layer increases and becomes dynamically dominant for high enough $R$, which can be considered the criterion that differentiates the low and high-velocity ratio regimes.

In addition, \citet{bucci_roughness-induced_2018} investigated a flow behind a single cylinder in a boundary layer, which represents JICF with the elevated jet nozzle in the limit of no jet flux ($R=0$). They observed shedding of vortical structures behind the obstacle, which indicates that the elevated jet pipe has potential to generate self-sustained shedding of hairpin vortices. \citet{acarlar_study1_1987} reported similar results for the wake of a single hemisphere.

Low-ratio jets can be used for film cooling, a process in which small flux of cold fluid from the orifice creates a thin layer over the surface, protecting it from the hot fluid in the free-stream \citep{jovanovic_effect_2008}. In this context hairpin vortices should be avoided, since their generation increases the mixing rate. A first step to understand and to control the hairpin generation is to determine the hairpin instability threshold and the location at which these structures are formed. The bifurcation that leads to periodic hairpin shedding in low R regime has been numerically investigated by \citet{ilak_bifurcation_2012} for a single crossflow Reynolds number $\Rey_D=495$. Based on the observation of self-sustained oscillations, the authors postulated that the hairpin shedding occurs through Hopf bifurcation. However, certain aspects of dynamics were not properly addressed due to large computational cost: they imposed the exit velocity profile at the jet orifice as a boundary condition instead of including the jet pipe inlet in the computational domain. \citet{peplinski_global_2015,peplinski_investigations_2015} compared the results obtained with two different computational grids (the first with an idealized jet as in \citet{ilak_bifurcation_2012} and the second with the inlet jet pipe directly included in the mesh) for $R=0.47$ (equivalently $R^*=1.5$) and $\Rey_D=495$, and observed that $R_{cr}$ may be significantly reduced when the pipe inlet is included. They also pointed out that numerical determination of critical velocity ratio $R_{cr}$ for hairpin shedding instability is difficult due to the sensitivity of simulations on the numerical method, grid size, and resolution. The influence of jet inlet pipe on the numerical simulations and the dependence of the hairpin dynamics on crossflow Reynolds numbers is the subject of ongoing work in KTH group \citep{chauvat_global_2017}.

Motivated by difficulties related to the numerical determination of the critical velocity ratio and possible influence of the jet inlet pipe  \citep{ilak_bifurcation_2012,peplinski_investigations_2015}, the main aim of this work is to quantitatively study the instability that gives rise to the self-sustained hairpin shedding. We investigate experimentally JICF varying both jet and free-stream speed and use spectral analysis to characterize the hairpin shedding (i.e. the study of the characteristic Strouhal number, global mode of hairpin vortices and bifurcation diagrams of the hairpin instability). We show that the hairpin amplitude can be described by Landau model, similar to other hydrodynamical oscillators such as wake behind bluff bodies \citep{mathis_benard-von_1984,goujon-durand_downstream_1994,wesfreid_global_1996}. Using this model, we determine the onset of the hairpin shedding instability for a specific parameter range. This is the first time when such a dynamical model was used to describe the hairpin shedding in JICF.

\section{Experimental set-up}
\begin{figure}

\begin{minipage}{0.49\textwidth} 
\hspace{0.4cm} \includegraphics[trim={0 10.35cm 0.25cm 9.4cm},clip,scale=0.27]{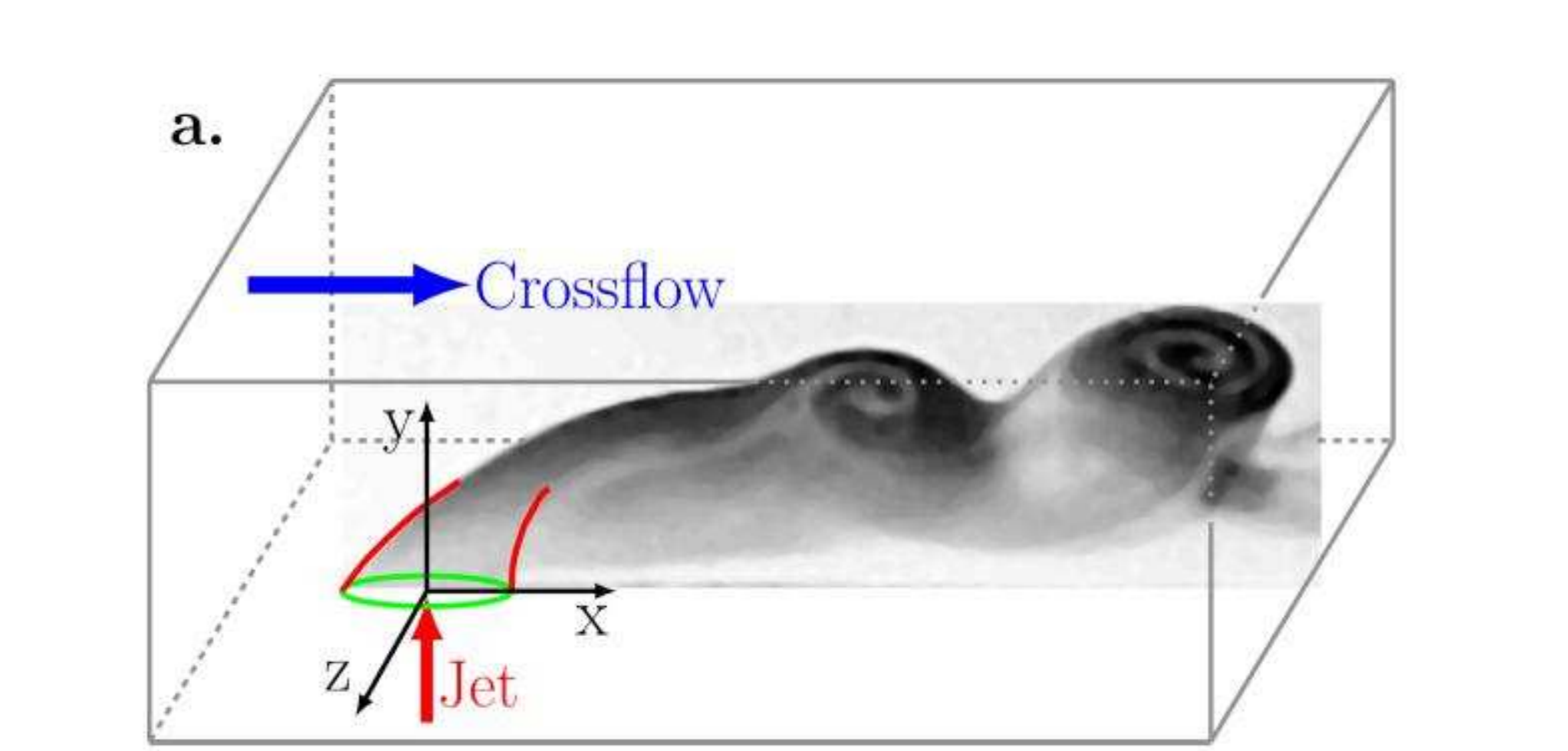}\\
\end{minipage}
\begin{minipage}{0.49\textwidth} 
\hspace{0.2cm} \includegraphics[trim={0 1.65cm 0.25 -0.5cm},scale=0.27]{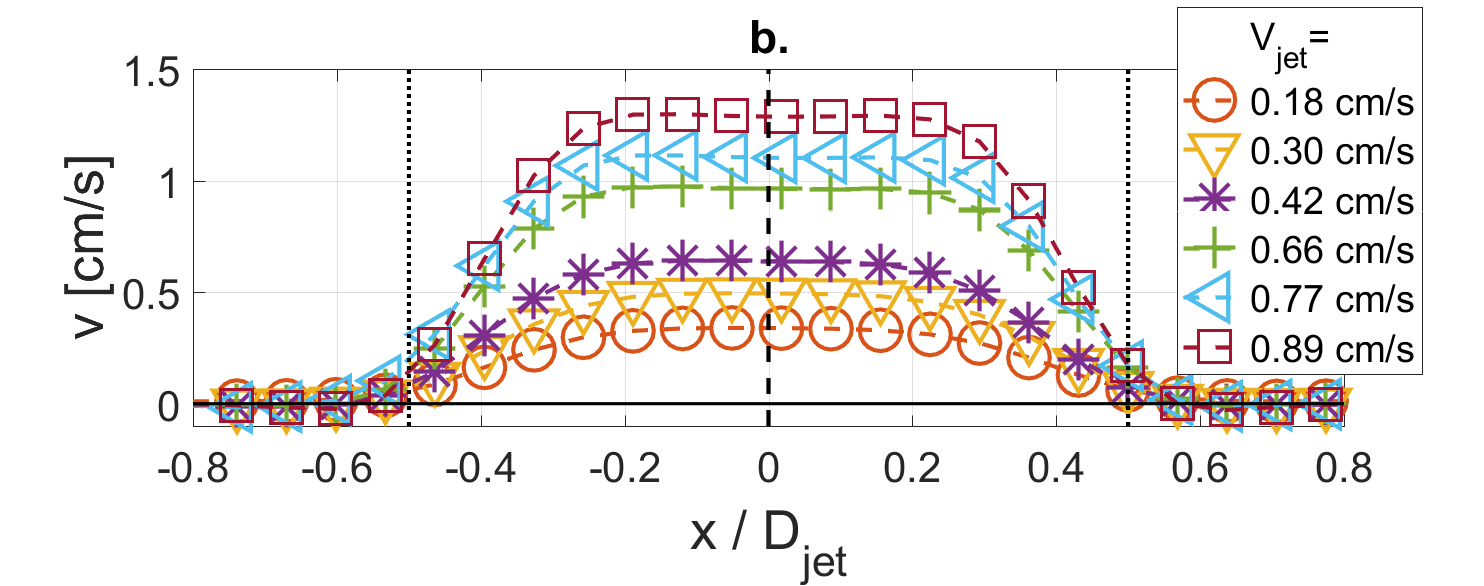}\\
\label{fig:CrossflowProfiles}   
\end{minipage} 
\caption{ a) Schematic representation of the experimental configuration indicating the streamwise ($x$), wall-normal ($y$), and spanwise ($z$) directions, respectively. Green ellipse represents the jet orifice. The scheme is supplemented with the picture of flow visualisation (streaklines) of the downstream near-field of JICF to illustrate the structure of hairpin vortices for $R=0.68$ and $\Rey_D=310$; b) time-averaged wall-normal velocity profiles measured above the jet orifice ($y/D_{jet}=0.07$) in the $z=0$ plane for different jet fluxes and in the absence of crossflow ($U_0 = 0$). Vertical black dashed line indicates the center of the jet, whereas vertical black dotted lines mark upstream and downstream edge of the jet orifice.}  
\label{fig:CrossflowAndJet}   
\begin{minipage}{0.49\textwidth}
\vspace{0.19cm}\includegraphics[trim={0 1.65cm 0.25 -0.5cm},scale=0.27]{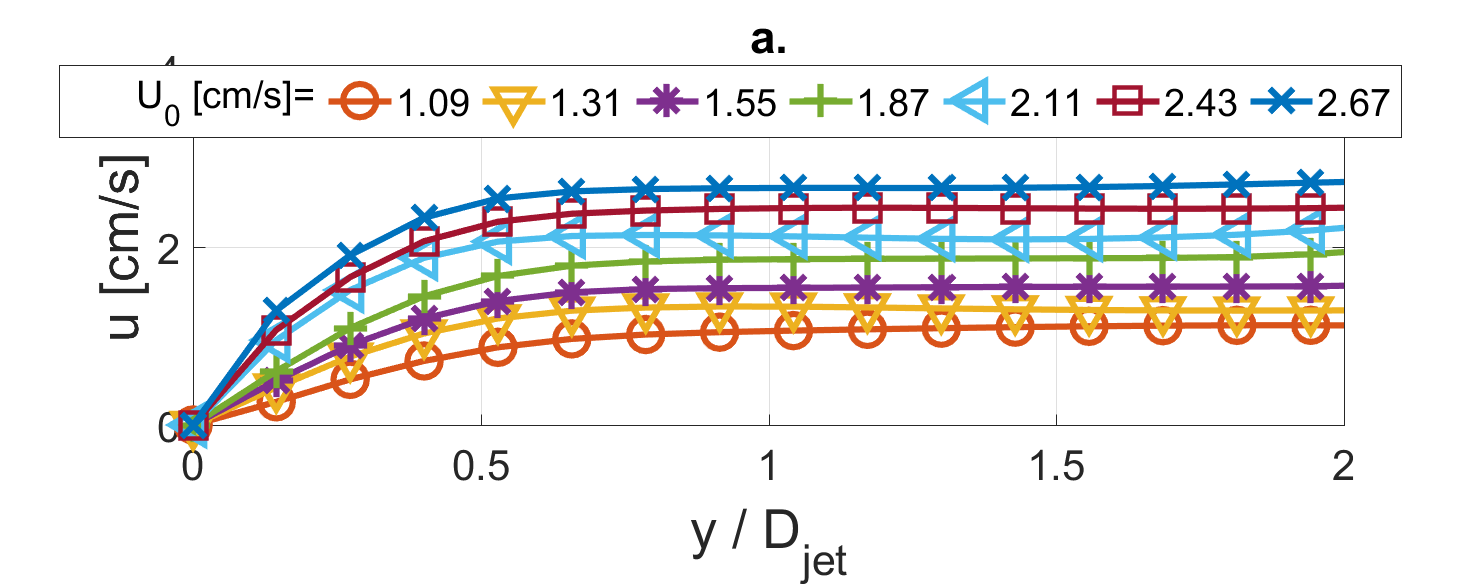}\\  
\end{minipage}
\begin{minipage}{0.49\textwidth} 
\hspace{0.05cm} \includegraphics[trim={0 1.65cm 0.25 -0.7cm},scale=0.27]{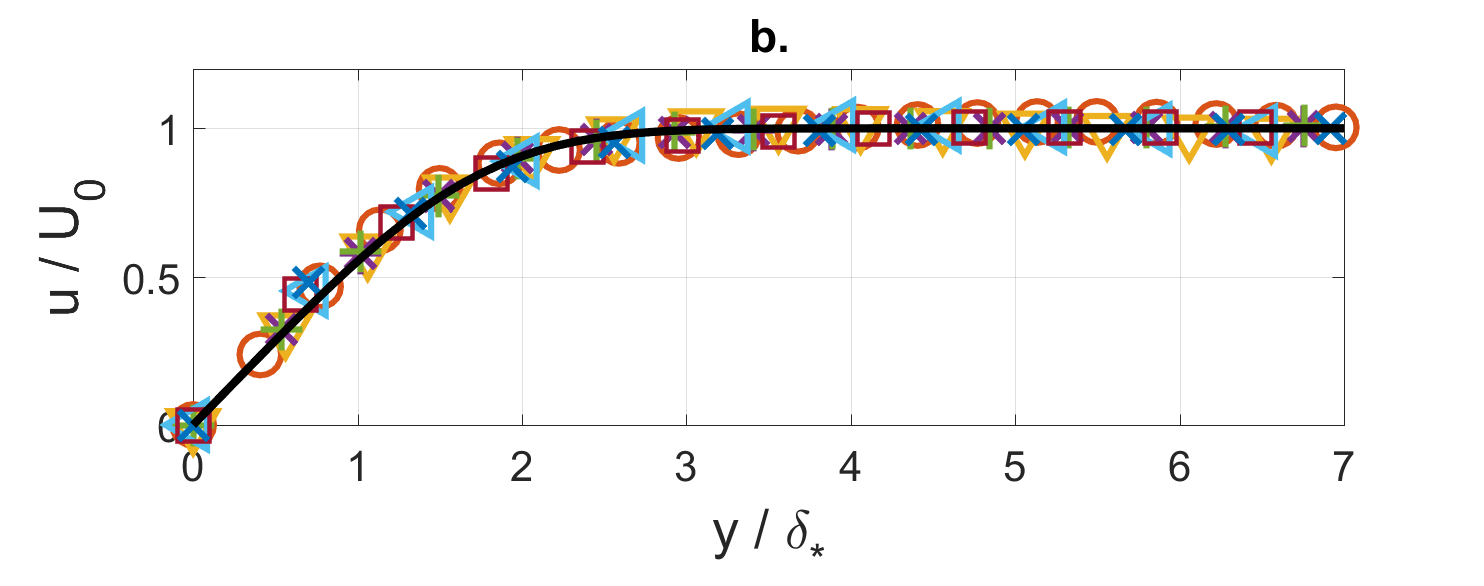}\\ 
\end{minipage} 
\caption{ a) Time-averaged streamwise velocity profiles $u(y)$ of the crossflow measured at $x/D_{jet}=-2.6$ in the symmetry plane ($z=0$) for different values of free-stream speeds $U_0$; b) the same data as in a) but with the wall-normal coordinate normalized with boundary layer displacement thickness $\delta_*$ (see Tab. \ref{tab:BLthickness}). Thick solid black curve represents the theoretical solution for laminar Blassius boundary layer.}  
\label{fig:CrossflowAndJetBlassius}   
\end{figure}

We use three non-dimensional parameters to characterise the JICF configuration: velocity ratio $R = V_{jet}/U_0$, crossflow Reynolds number $Re_{D}=U_0D_{jet}/\nu$, and jet Reynolds number $Re_{jet}=V_{jet}D_{jet}/\nu=R \cdot Re_{D}$, where $U_0$, $V_{jet}$, $D_{jet}$, and $\nu$ are free-stream and jet bulk velocities, jet diameter and kinematic viscosity of the fluid, respectively. All quantities are normalized with appropriate combination of $D_{jet}$ and $U_0$. The experimental configuration along with the typical vortical structures in the flow are shown in Fig. \ref{fig:CrossflowAndJet}a. We place the origin of the coordinate system $(0,0,0)$ at the center of the jet orifice and denote by $x,y,z$ the streamwise (direction of the free-stream), wall-normal (vertical, direction of the jet) and spanwise directions. Experiments were carried out in a close loop channel at Warsaw University of Technology, with water at room temperature as the working fluid for both jet and crossflow. The experimental set-up has a horizontal test section made of plexiglass to provide optical access. The test section has rectangular cross-section of $15.0$ cm width and $10.0$ cm height, and length of the channel is $150.0$ cm. The crossflow is induced by gravity, using a constant-level tank located above the test section to provide a constant pressure gradient (see also Fig.1 in \citet{klotz_experimental_2014}). Before entering the test section, the crossflow is conditioned by two honeycomb screens and a 4:1 contraction nozzle. The bulk crossflow velocity is controlled by a valve and measured with tensometric weight with $\pm0.5\%$ accuracy.

\begin{table}
  \begin{center}
\def~{\hphantom{0}}
  \begin{tabular}{cccccccccccc}
       $U_0$ [cm/s]              & $1.09$     &   $1.31$ & $1.55$  & $1.87$ & $2.11$ & $2.43$ & $2.67$& & &\\
       $Re_D = U_0 D_{jet}/ \nu$ & $260$      &   $310$  & $370$   & $450$  & $510$  & $590$  & $640$ \\  
       $\delta_* / D_{jet}$ (for $x/D_{jet}=-2.6$)     & $0.35$     &   $0.26$ & $0.27$  & $0.27$ & $0.20$ & $0.22$ & $0.21$& & &\\
	   $\delta_{**} / D_{jet}$ (for $x/D_{jet}=-2.6$)  & $0.14$     &   $0.09$ & $0.10$  & $0.10$ & $0.08$ & $0.09$ & $0.08$& & &\\  
  \end{tabular}
  \caption{First and second rows indicate all investigated crossflow speeds $U_0$ and corresponding crossflow Reynolds numbers $\Rey_D$, respectively. Third and fourth rows show displacement thickness $\delta_*$ and momentum loss thickness $\delta_{**}$ of lower boundary layer formed above the floor of the test section at $x/D_{jet}=-2.6$ location. These characteristic quantities are calculated by trapezoidal integration from mean streamwise velocity profiles shown in Fig. \ref{fig:CrossflowAndJetBlassius}.}
  \label{tab:BLthickness}

  \end{center}
\end{table}

The jet is injected normal to the wall into the crossflow through a circular exhaust with an internal diameter $D_{jet} = 2.17$ cm, mounted flush with the test section floor. The jet plenum chamber is supplied with the fluid by an additional pump through a spiral tube with several holes drilled along its length. The jet flow is then conditioned by a porous material and a 16:1 contraction nozzle, and is controlled by a precise needle valve. The bulk velocity of the jet is measured by a calibrated Kobold rotameter with $\pm1.2\%$ accuracy.

We measure velocity fields with a 2D Particle Image Velocimetry (PIV) set-up that consists of a Litron Nano L200-15 laser (double-headed, 532 nm light, 1200 mJ energy per pulse) and Imager sCMOS camera (16-bit, 2560 x 2160 pix). We acquire single-frame image sequences and cross-correlate two consecutive images using Davis 8.1 Lavision software using 32 x 32 pixel interrogation windows with 50\% overlap. To retain the time correlation between two snapshots, for each measurement we use a frequency from 7 Hz to 30 Hz (depending on $U_0$) that corresponds to about $0.06$ advective time units ($t_{adv} = D_{jet} / U_0$), unless otherwise stated. For all PIV measurements presented in this paper the laser sheet, of about $1.5$ mm thickness, is aligned with $z=0$ plane. 

First, we measure the wall-normal velocity component of the jet in the absence of the crossflow ($U_0 = 0$). For this, we acquire 100 images for each jet flux, adjusting the acquisition frequency such that the time between two consecutive frames is about $0.03$ advective time units (in this case defined as $t_{adv}=D_{jet}/V_{jet}$). In Fig. \ref{fig:CrossflowAndJet}b we present a time-averaged wall-normal velocity component measured at $y/D_{jet}=0.07$ for different jet fluxes. Black vertical dashed and dotted lines mark the jet center and edges of the jet orifice, respectively. Temporal fluctuations of $V_{jet}$ are lower than 0.8\%.

Next, we use PIV to measure the streamwise velocity component of the crossflow at $x/D_{jet}=-2.6$ and as a function of the wall-normal coordinate. For each crossflow speed we measured 10 sets of data, each containing 660 images. The mean velocity profiles in Fig. \ref{fig:CrossflowAndJetBlassius}a are averaged over time and over all sets. Then we calibrate the bulk crossflow speed into the free-stream velocity $U_0$. We also estimate the turbulence intensity level by calculating standard deviation (STD) of the streamwise velocity component; spatial/temporal fluctuations of free-stream velocity (in the central part of the test section) are lower than 2.3\%/1.8\%, respectively, and temporal velocity fluctuations of lower boundary layer are lower than 3.9\%. The wall-normal velocity fluctuations of the cross-flow are $\simeq 0.8 \%$ of $U_0$ for $U_0$ lower than $2.11$ cm/s ($\Rey_{D}=510$) and start to increase up to $\simeq 3.1 \%$ of $U_0$ for $U_0=2.67$ cm/s ($\Rey_{D}=640$). The position of the lower bounding wall and the distance to the first y-position was deduced from the raw images. Displacement thickness $\delta_*$ and momentum loss thickness $\delta_{**}$ of lower boundary layer are calculated at $x/D_{jet}=-2.6$ by trapezoidal integration and are shown in Tab. \ref{tab:BLthickness}. The resulting shape factor $H = \delta_*/\delta_{**}$ is equal to $2.6 \pm 0.2$, close to the value of laminar Blassius boundary layer. In Fig. \ref{fig:CrossflowAndJetBlassius}b we plot the time-averaged streamwise velocity profiles of the crossflow at $x/D_{jet}=-2.6$ using the free-stream speed $U_0$ to normalize the streamwise velocity component and boundary layer displacement thickness $\delta_*$ to normalize the wall-normal coordinate. For all considered crossflow velocities our measurements compare well with the streamwise velocity profile of laminar Blassius boundary layer, shown as thick black curve in Fig. \ref{fig:CrossflowAndJetBlassius}b.

\section{Experimental results}
\begin{figure}
\hspace{0.3cm}\includegraphics[trim={0 0cm 0 -0.5cm},scale=0.30]{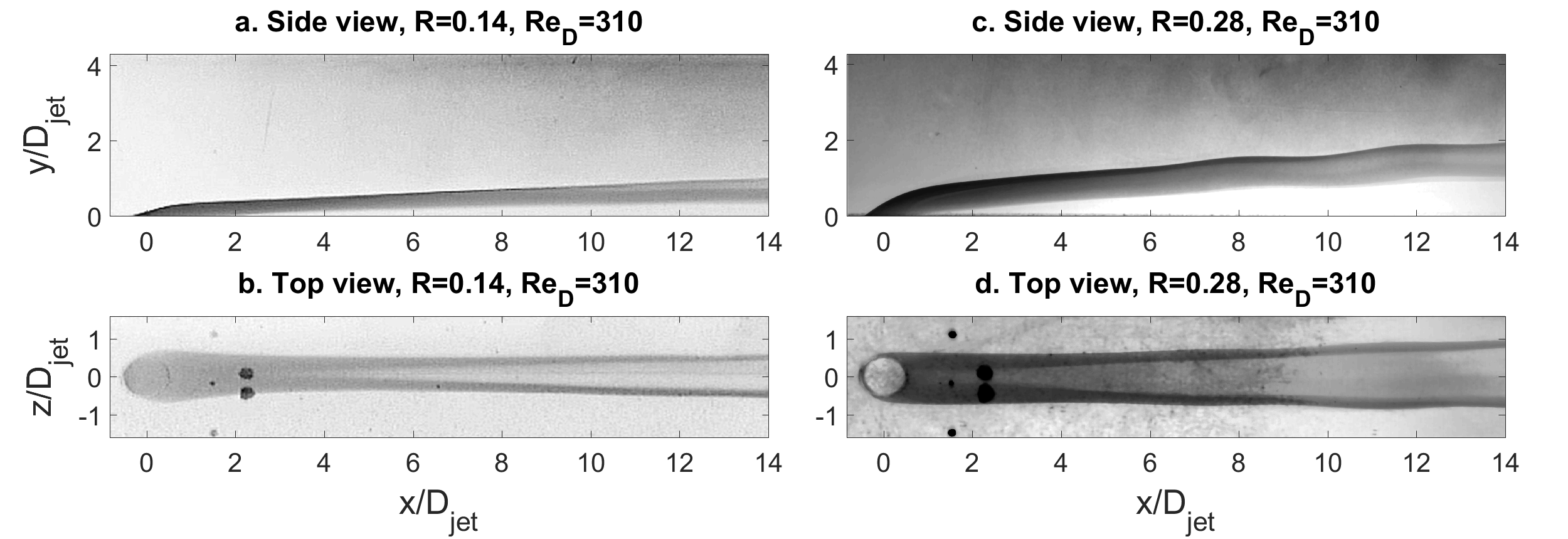}\\
\\
\includegraphics[trim={-1.1cm 1.15cm 0 -0.5cm},scale=0.30]{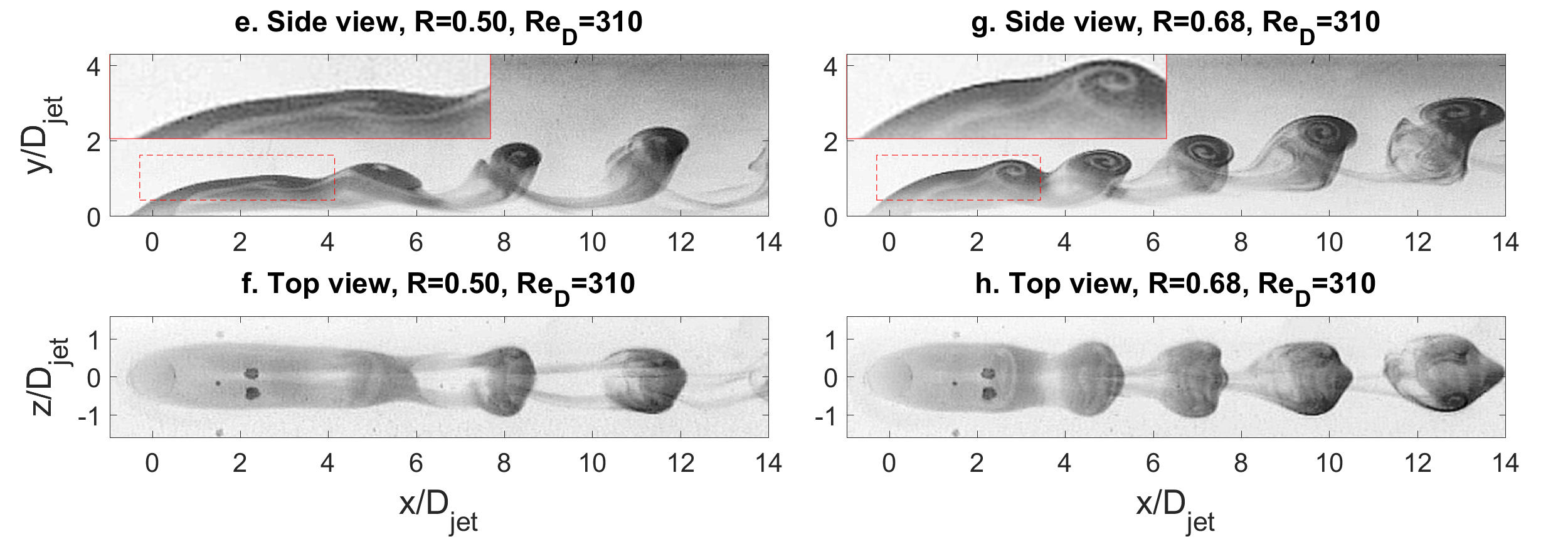}\\
\caption{ Side (a,c,e,g) and top (b,d,f,h) view of flow visualisations (streaklines) for $\Rey_D=310$ to illustrate how the flow structure changes when velocity ratio $R$ is increased from $R=0.14$ to $R=0.68$. Insets at top-left corner in e) and g) present at double magnification the region marked by dashed red rectangles in the corresponding subplots.} 
\label{fig:Visu}    
\hspace{-0.50cm} \includegraphics[trim={0 1.95cm 0.25 -0.7cm},scale=0.27]{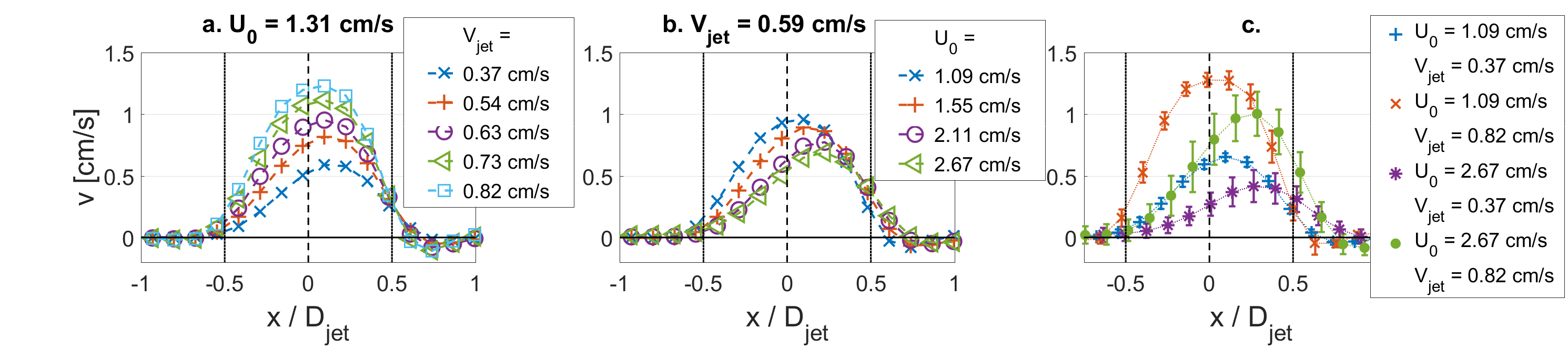}\\ 
\caption{ Illustration of the modification of the wall-normal jet velocity profile $v(x)$ due to to interaction with $U_0$: a) $v(x)$ for different bulk jet speeds and for fixed $U_0=1.31$ cm/s; b) $v(x)$ for different $U_0$ and for fixed $V_{jet}=0.59$ cm/s; c) $v(x)$ for the minimal ($V_{jet}=0.37$ cm/s) and maximal ($V_{jet}=0.82$ cm/s) jet bulk speed and for the minimal ($U_0=1.09$ cm/s) and maximal ($U_0=2.67$ cm/s) crossflow speed. Error bars at each $x$ and each combination of $V_{jet}$ and $U_0$ indicate temporal fluctuations of $V_{jet(t)}$, calculated as a standard deviation of instantaneous wall-normal jet velocity $V_{jet}(t)$. In order to increase readability, the values of STD are premultiplied by a factor of two. \vspace{0.4cm}}  
\label{fig:JetModificationDueCross}
\end{figure}

\begin{figure}
\hspace{0.0cm} \includegraphics[trim={0 1.65cm 0 -0.35cm},scale=0.32]{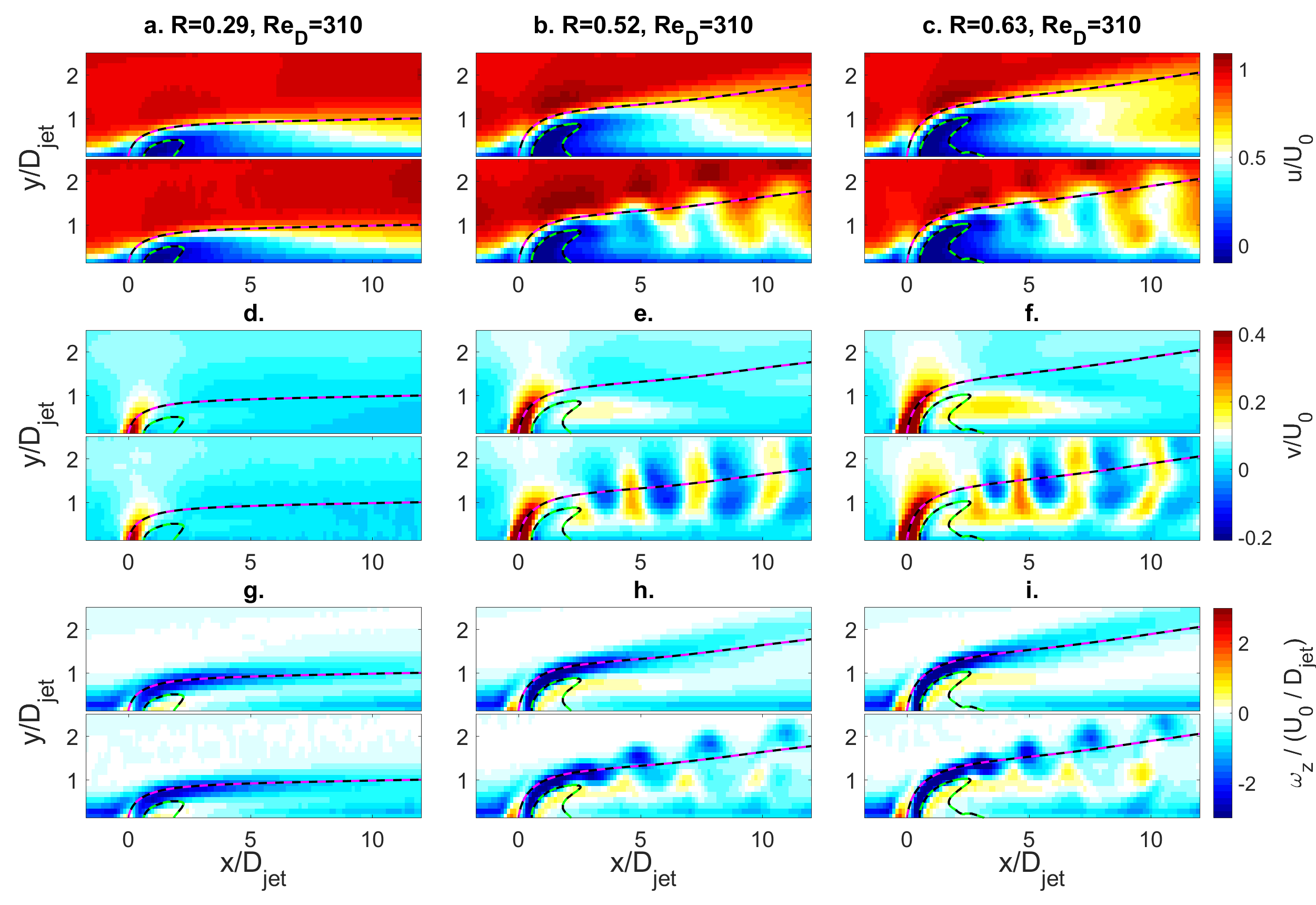}\\
\caption{ Results from 2D PIV measurements at $z=0$ and for $\Rey_D=310$ showing the streamwise $u$ (a,b,c) and wall-normal $v$ (d,e,f) velocity components and the spanwise vorticity $\omega_z$ (g,h,i). First (a,d,g), second (b,e,h), and third (c,f,i) column corresponds to $R=0.29$, $R=0.52$, and $R=0.63$, respectively. Each subplot is composed of the time-averaged (top) and instantaneous (bottom) fields. On each field time-averaged centerline trajectory (magenta-black dashed line) and downstream recirculation zone (green-black dashed line) are superimposed.} 
\label{fig:PIV1} 
\hspace{0.0cm} \includegraphics[trim={0 1.65cm 0 -0.35cm},scale=0.32]{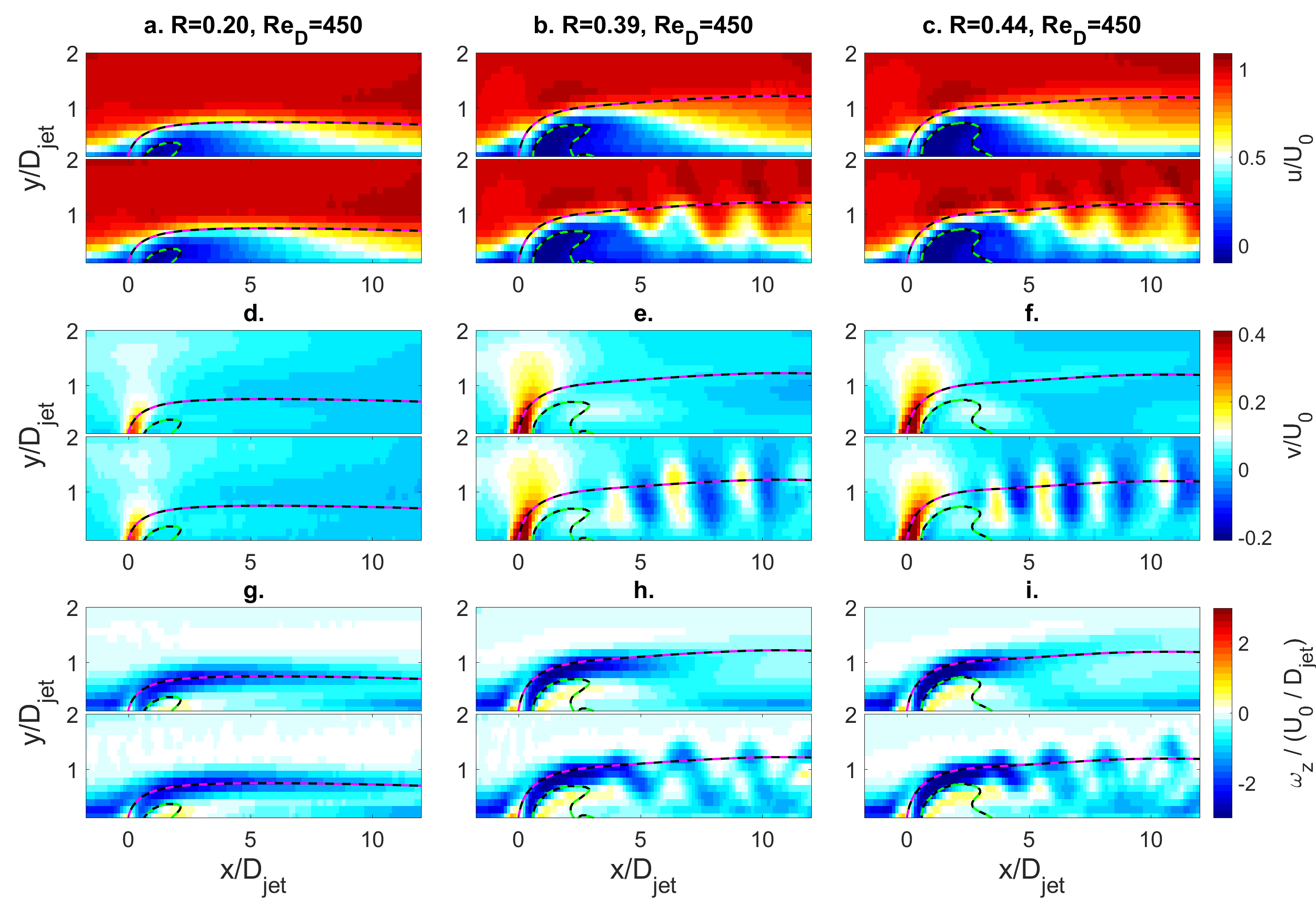}\\ 
\caption{ The same as for Fig. \ref{fig:PIV1} but for $\Rey_D=450$. First (a,d,g), second (b,e,h), and third (c,f,i) column corresponds to $R=0.20$, $R=0.39$, and $R=0.44$.} 
\label{fig:PIV2} 

\end{figure}
\begin{figure}
\hspace{0.0cm} \includegraphics[trim={0 1.65cm 0 -0.35cm},scale=0.32]{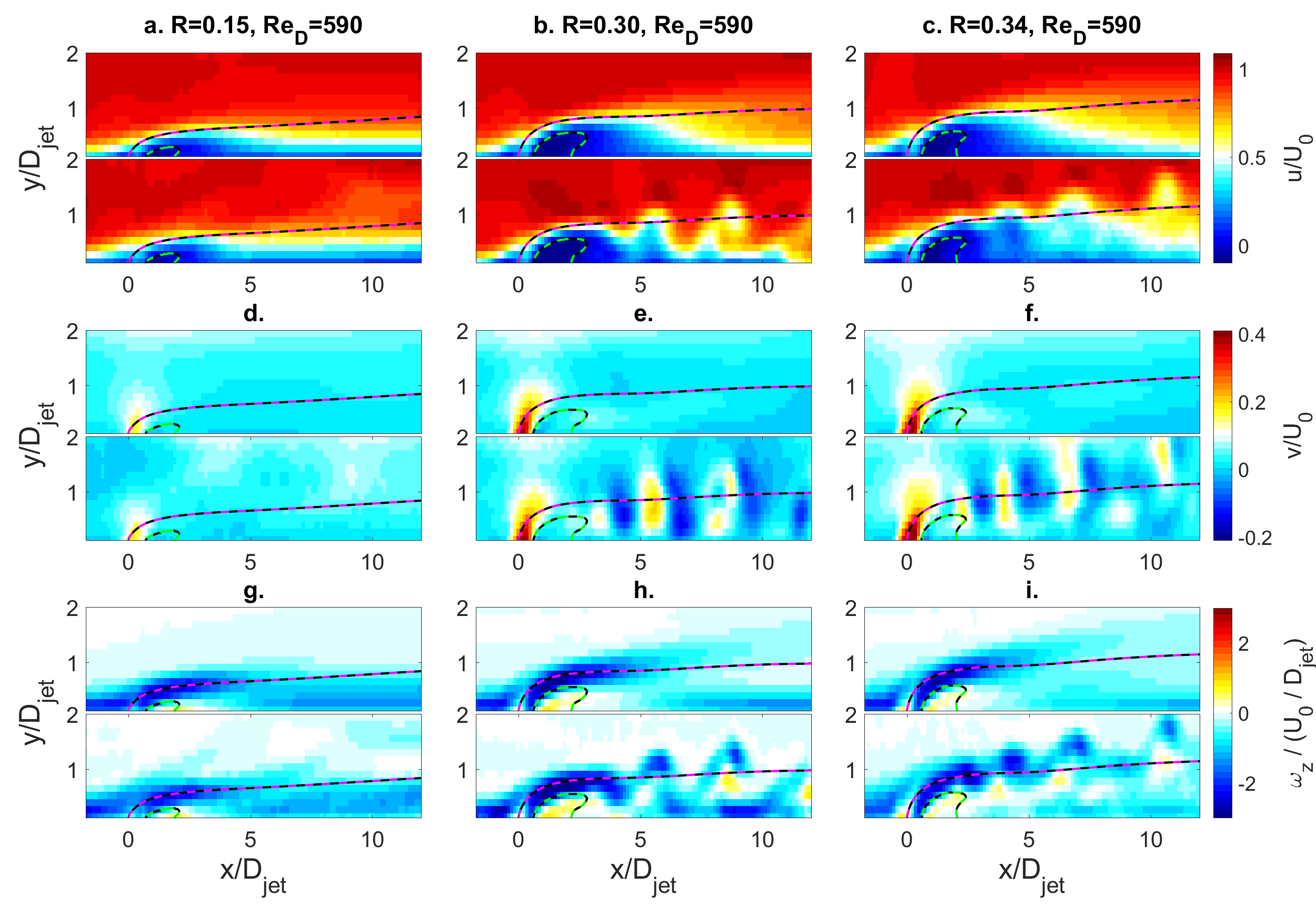}\\ 
\caption{ The same as for Fig. \ref{fig:PIV1} but for $\Rey_D=590$. First (a,d,g), second (b,e,h), and third (c,f,i) column corresponds to $R=0.15$, $R=0.30$, and $R=0.34$.  \vspace{0.4cm}}
\label{fig:PIV3} 
\end{figure}

\begin{figure}
\hspace{0.0cm} \includegraphics[trim={0 1.65cm 0 0.15cm},scale=0.32]{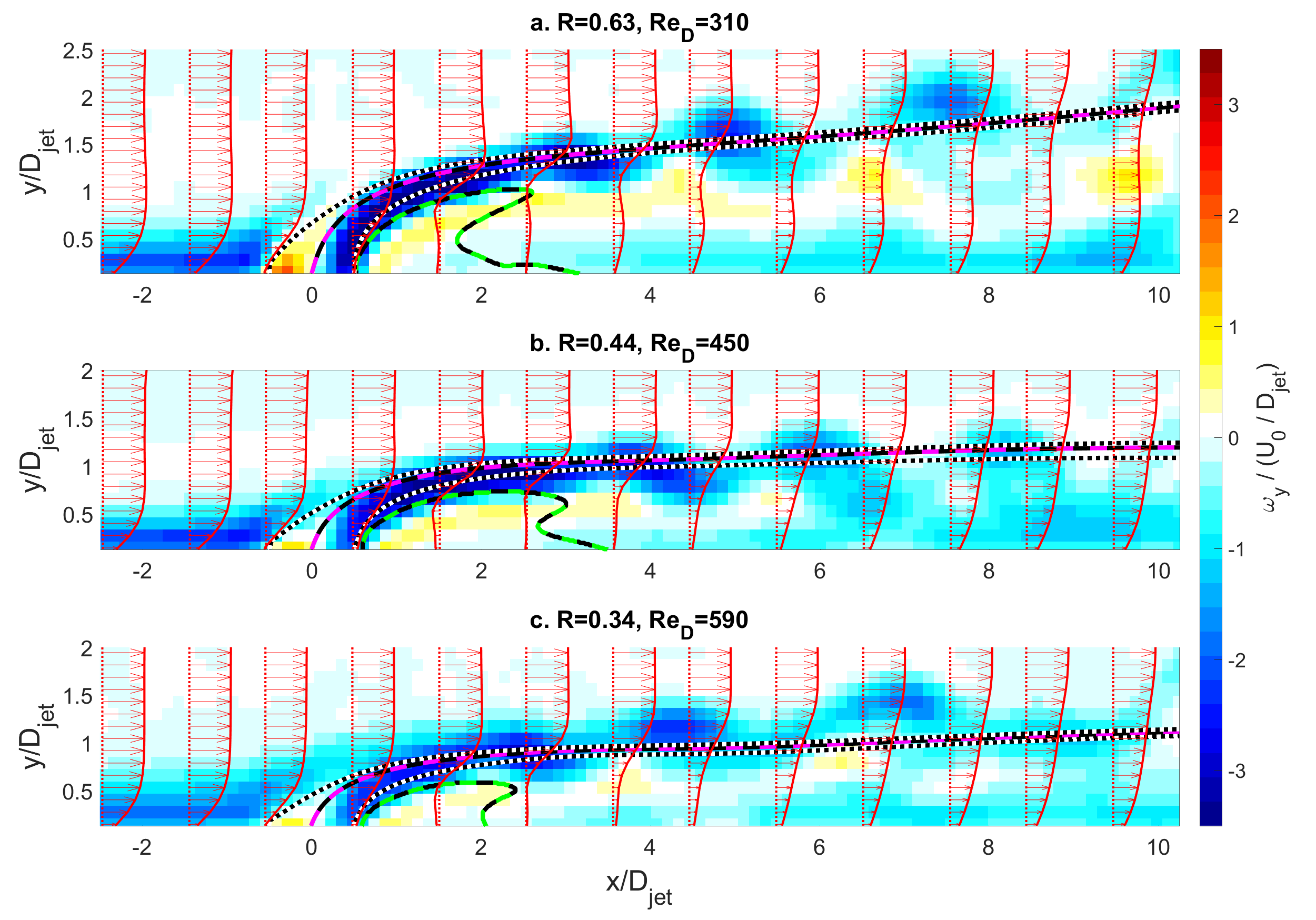}\\ 
\caption{ Instantaneous spanwise vorticity fields illustrating hairpin vortices for: a) $R=0.63$ and $\Rey_D=310$; b) for $R=0.44$ and $\Rey_D=450$; c) for $R=0.34$ and $\Rey_D=590$. Time-averaged streamwise velocity profiles for different streamwise locations are also plotted (red vectors) to illustrate the spatial evolution of boundary layer and its modification due to the operating jet. Magenta-black dashed curve and green-black dashed curve represents time-averaged centerline trajectory and downstream recirculation zone, respectively. In addition, the trajectories emanating from upstream ($x/D_{jet} = -0.5$) and downstream jet shear layer ($x/D_{jet} = 0.5$) are also plotted by black-white dotted curves. \vspace{0.4cm}}
\label{fig:StreamVelocityEvolution} 
\end{figure}

\begin{figure}
\hspace{0.0cm} \includegraphics[trim={0 1.65cm 0 0.15cm},scale=0.32]{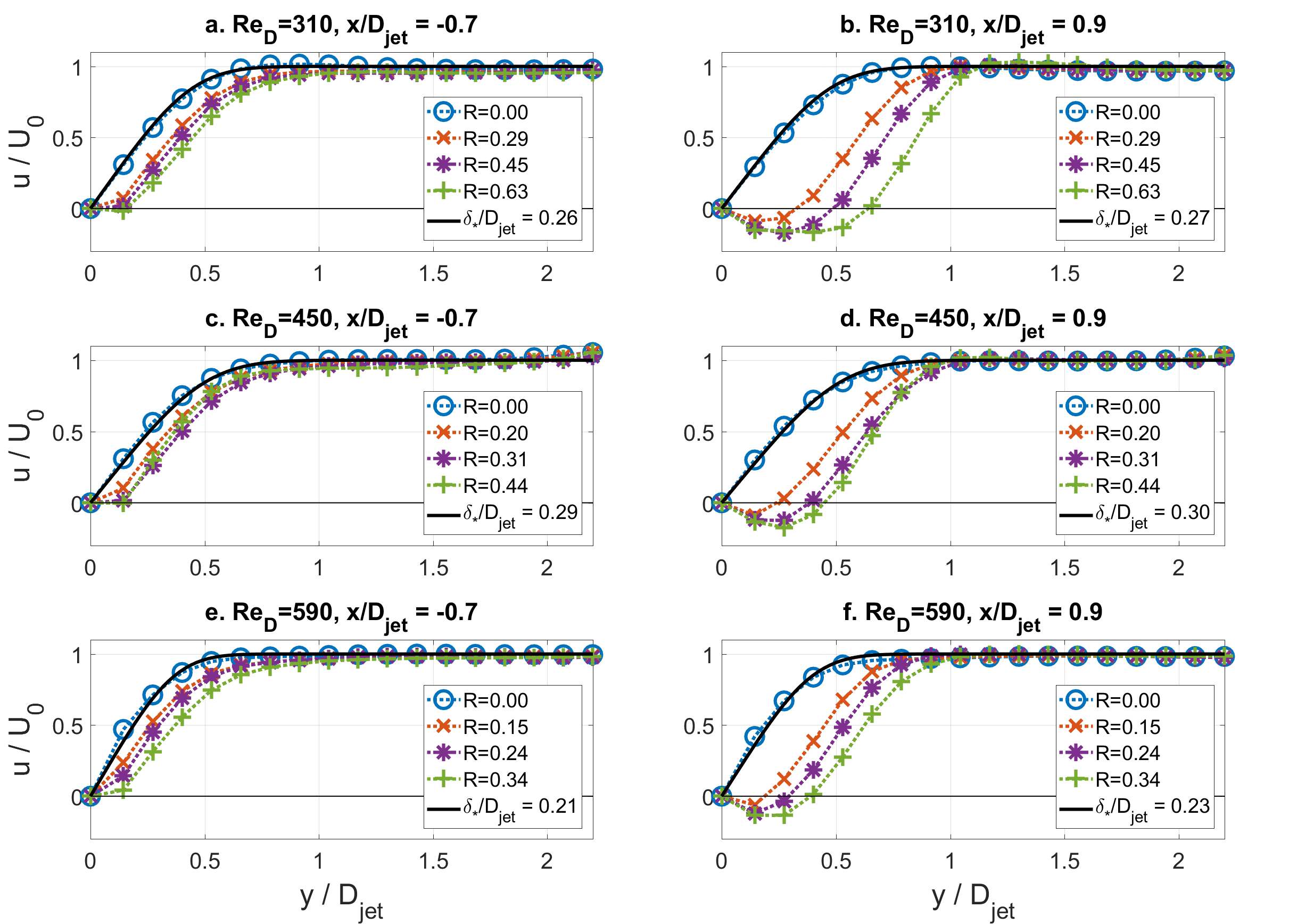}\\ 
\caption{ Time-averaged streamwise velocity profiles $u(y)$ along the wall-normal direction measured at $x/D_{jet}=-0.7$ (first column) and $x/D_{jet}=0.9$ (second column). First, second, and third row correspond to $\Rey_D=310$, $\Rey_D=450$, and $\Rey_D=590$, respectively. For each velocity profile the corresponding value of velocity ratio $R$ is indicated in the legend. The streamwise velocity profile $U(y)$ of the base flow without jet perturbation ($R=0$, blue circles) is also compared with the theoretical solution of laminar Blassius boundary layer (thick black curve). The displacement thickness of boundary layer ($\delta_*/D_{jet}$) for each case is indicated in the legend. \vspace{0.4cm}} 
\label{fig:StreamVelocityEvolutionNearJet} 
\end{figure}

To illustrate qualitatively the flow dynamics of JICF in low-velocity-ratio regime, we first present flow visualisations (streaklines) using fluorescein dye excited by a point source of visible light. The concentrated fluorescent colourant was injected from a pressurized container into the jet bulk flow before the jet plenum chamber. This results in a uniform fluoresceine concentration across the entire jet cross-section at $y=0$. Fig. \ref{fig:Visu} illustrates for $\Rey_D=310$ how the flow structure changes when $R$ is increased from $R=0.14$ to $R=0.68$. The top and bottom rows correspond to the side and top views, respectively. Both views are captured instantaneously with a single Nikon D610 camera (with 6016 x 4016 pix matrix) and a mirror inclined at 45 degrees with respect to horizontal plane. For high enough $R$ the downstream near-field of JICF is dominated by a distinct periodic shedding of hairpin vortices (Fig. \ref{fig:Visu}e,f,g,h). Insets in the top-left corner of subplots e) and g) of Fig. \ref{fig:Visu} present at double magnification the region marked by dashed red rectangle in the corresponding subplots. The spirals formed within hairpin heads in Fig. \ref{fig:Visu}g rotate clockwise, which corresponds to negative spanwise vorticity. Similar hairpin vortices were also reported by \citet[][]{lim_development_2001,camussi_experimental_2002,ilak_bifurcation_2012}. In contrast, for low $R$ no distinct self-sustained hairpin shedding can be observed in the downstream near-field (Fig. \ref{fig:Visu}a,b,c,d). We note that the transition to hairpin shedding in JICF in low-velocity-ratio regime is analogous to the transition reported for the wake behind three-dimensional bluff bodies, such as sphere \citep{johnson_flow_1999,gumowski_transition_2008}, disk \citep{bobinski_instabilities_2014} and cube \citep{klotz_experimental_2014}, with similar structure of shed hairpins and with a gentle kinking of two trails of counter-rotating vortices prior to self-sustained shedding of hairpins (see at $x/D_{jet}>8$ in Fig. \ref{fig:Visu}c,d). In addition, \citet{peplinski_global_2015} studied numerically JICF for $R\in(0.31,0.50)$ and $\Rey_D=495$, and observed similar kinking of two-counter rotating vortices for $R$ below the threshold of hairpin shedding. They proposed non-modal transient growth of the perturbation (see \citeauthor{schmid_stability_2001}; \citeyear{schmid_stability_2001} for theoretical description and \citeauthor{klotz_experiments_2017}; \citeyear{klotz_experiments_2017} for experimental verification) as a mechanism leading to these oscillations.  

To evaluate quantitatively JICF behaviour, we acquire 3 sequences of 660 images for each $U_0$ (see Tab. \ref{tab:BLthickness}) and $V_{jet}<1$ cm/s, which corresponds to $\Rey_D=U_0\,D_{jet}/\nu \in (260,640)$ and $R=V_{jet}/U_0\in(0.14,0.75)$. The precise range of $R$ is different for each $U_0$. Flow visualisations (Fig. \ref{fig:Visu}) and PIV measurements presented below were carried out in different runs. The measurement region covers $x/D_{jet}\in (-3.95,18.33)$ and $y/D_{jet}\in (0.14, 4.12)$, with spatial resolution of $0.13 D_{jet}$. First, we evaluate a modification of the jet velocity profile when $U_0\neq 0$. In Fig. \ref{fig:JetModificationDueCross} we present a time-averaged wall-normal jet velocity component as a function of $x$, measured at $y/D_{jet}=0.26$. For fixed $U_0= 1.31$ cm/s (Fig. \ref{fig:JetModificationDueCross}a) the shape of the jet profile remains relatively unchanged and the maximal jet velocity increases proportionally with $V_{jet}$. In contrast, for fixed $V_{jet}=0.59$ cm/s (Fig. \ref{fig:JetModificationDueCross}b) the maximal vertical jet velocity decreases and shift toward the trailing edge of the jet when $U_0$ is increased. In Fig. \ref{fig:JetModificationDueCross}c we show the time-averaged jet profiles for all possible combinations of lowest/highest values of jet ($V_{jet}=0.37$ and $0.82$  cm/s) and crossflow ($U_0 = 1.09$ and $2.67$ cm/s) speed. These four cases span the entire parameter space investigated in this paper. To estimate temporal fluctuations of jet profile when $U_0\neq0$, we calculate the standard deviation of $V_{jet}(t)$ for each $x$ location. The resulting values, premultiplied by a factor of two, are plotted as error bars in Fig. \ref{fig:JetModificationDueCross}c. For low $V_{jet}$ and $U_0$, the jet profile is almost completely stationary and when either $V_{jet}$ or $U_0$ are increased, temporal fluctuations of the jet profile also increase. Even for the lowest $V_{jet}$ and the highest $U_0$ we do not observe any significant flow from the crossflow into the jet orifice near the leading edge.

The first (a,b,c), second (d,e,f) and third (g,h,i) rows in Fig. \ref{fig:PIV1} represent the streamwise ($u$) and wall-normal ($v$) velocity components and spanwise vorticity ($\omega_z$), left to right. The measurements were carried out for $\Rey_D=310$ and for $R=0.29,0.52,0.63$. Each subplot is composed of the time-averaged (top) and instantaneous (bottom) flow fields. The spatial evolution of $u$, $v$, and $\omega_z$ of the flow field in the first column ($R=0.29$, Fig. \ref{fig:PIV1}a,d,g) is characterized by slow variation along the streamwise direction and no self-sustained oscillations can be observed. In contrast, for $R=0.52$ (Fig. \ref{fig:PIV1}b,e,h) and $R=0.63$ (Fig. \ref{fig:PIV1}c,f,i) distinct periodic fluctuations of $u$, $v$, and $\omega_z$ can be observed in the downstream near field of JICF ($0<x/D_{jet}<10$), which is a signature of shedding of hairpin vortices. In Fig. \ref{fig:PIV2}-\ref{fig:PIV3} a similar transition is presented for $\Rey_D=450$ and $\Rey_D=590$. However, when the crossflow Reynolds number is increased to $\Rey_D=590$, the coherence of the hairpin structure slightly decreases due to the increase of base flow fluctuations. In Fig. \ref{fig:PIV1}-\ref{fig:PIV3} we superimpose the centerline trajectory (the streamline starting at $x=0,y=0$ represented by magenta-black dashed curve) derived from time-averaged velocity fields and the downstream recirculation zone (green-black dashed curve at which $<u>_t=0$). This recirculation zone is present also as an instantaneous flow feature, in agreement with numerical simulations of \citet{schlatter_self-sustained_2011}. The downstream near-field ($0<x/D_{jet}<10$) of the JICF is dominated by negative spanwise vorticity $\omega_z<0$, which is of the same sign as the vorticity emanating from the downstream jet shear layer. This observation holds for the entire parameter space range considered here and is consistent with the numerical results of \citet{sau_dynamics_2008,ilak_bifurcation_2012}.

Both time-averaged (top) and instantaneous (bottom) fields in Fig. \ref{fig:PIV1}-\ref{fig:PIV3} are similar in the vicinity of the jet orifice. We did not observe any specific dynamics of the separated flow in the jet pipe, as reported by \citet{kelso_horseshoe_1995,bidan_steady_2013}, which may be due to the fact that our value of crossflow Reynolds number is lower when compared to \citet[][]{kelso_horseshoe_1995} ($\Rey_D > 1200$) and \citet[][]{bidan_steady_2013} ($\Rey_D \simeq 2700$). The only observed influence of the crossflow on the jet close to the jet orifice in our study is the deflection of the jet profile such that the maximal wall-normal velocity component of the jet is reached slightly downstream with respect to the jet center, as shown in Fig. \ref{fig:JetModificationDueCross}. Moreover, we did not measure any roll-up of the upstream jet shear layer as reported by \citet{getsinger_structural_2014,gevorkyan_influence_2018} who considered much higher jet Reynolds number ($\Rey_{jet}=1900$).

In order to better understand the structure of hairpin vortices we plot the spatial evolution of the time-averaged streamwise velocity profiles superimposed on the instantaneous spanwise vorticity field for $R\neq 0$ and for $\Rey_D=310,450,590$ (Fig. \ref{fig:StreamVelocityEvolution}). We also superimpose time-averaged recirculation zone (green-black dashed curve) and three different trajectories starting at $x/D_{jet}=0$ (magenta-black dashed curve) and at $x/D_{jet}=-0.5,0.5$ (black-white dotted curves). The boundary layer separates downstream from the issuing jet and forms a distinct recirculation zone. Further downstream, the recirculation zone hairpin vortices are formed. These vortical structures are composed mainly of the negative spanwise vorticity ($\omega_z<0$) located close to the centerline trajectory. However, the signature of hairpin vortices also composes of some weaker patches of positive spanwise vorticity ($\omega_z>0$) located below the centerline trajectory, which originates from the downstream wake formed by the issuing jet. To evaluate the influence of the jet perturbation on the boundary layer, we plot the time-averaged streamwise velocity profiles along the wall-normal direction and in the vicinity of the upstream ($x/D_{jet}=-0.6$, first column in Fig. \ref{fig:StreamVelocityEvolutionNearJet}) and the downstream ($x/D_{jet}=0.9$, second column in Fig. \ref{fig:StreamVelocityEvolutionNearJet}) edges of the jet orifice for three different crossflow Reynolds numbers (first row: $\Rey_D=310$, second row: $\Rey_D=450$, and third row: $\Rey_D=590$) in Fig. \ref{fig:StreamVelocityEvolutionNearJet}. On each subplot of Fig. \ref{fig:StreamVelocityEvolutionNearJet} the undisturbed laminar boundary layer, without jet perturbation ($R=0$), is represented by blue circles. These undisturbed velocity profiles are also compared with theoretical solution of the laminar Blassius boundary layer (thick black curves). The wall-normal coordinate of the theoretical profiles is re-scaled with $\delta_* / D_{jet}$, where $\delta_*$ is displacement thickness calculated for undisturbed streamwise velocity profiles measured experimentally. In addition, for each combination of spatial location and crossflow Reynolds number, we plot the streamwise velocity profile for three different velocity ratios, the specific values of which can be found in the legend. In all cases the boundary layer thickness, as well as the size of the downstream recirculation zone in the wall-normal direction, increase monotonically with $R$. 

\section{Spectral analysis}

\begin{figure}
\hspace{-0.2cm} \includegraphics[trim={0 1.65cm 0 -1.5cm},scale=0.32]{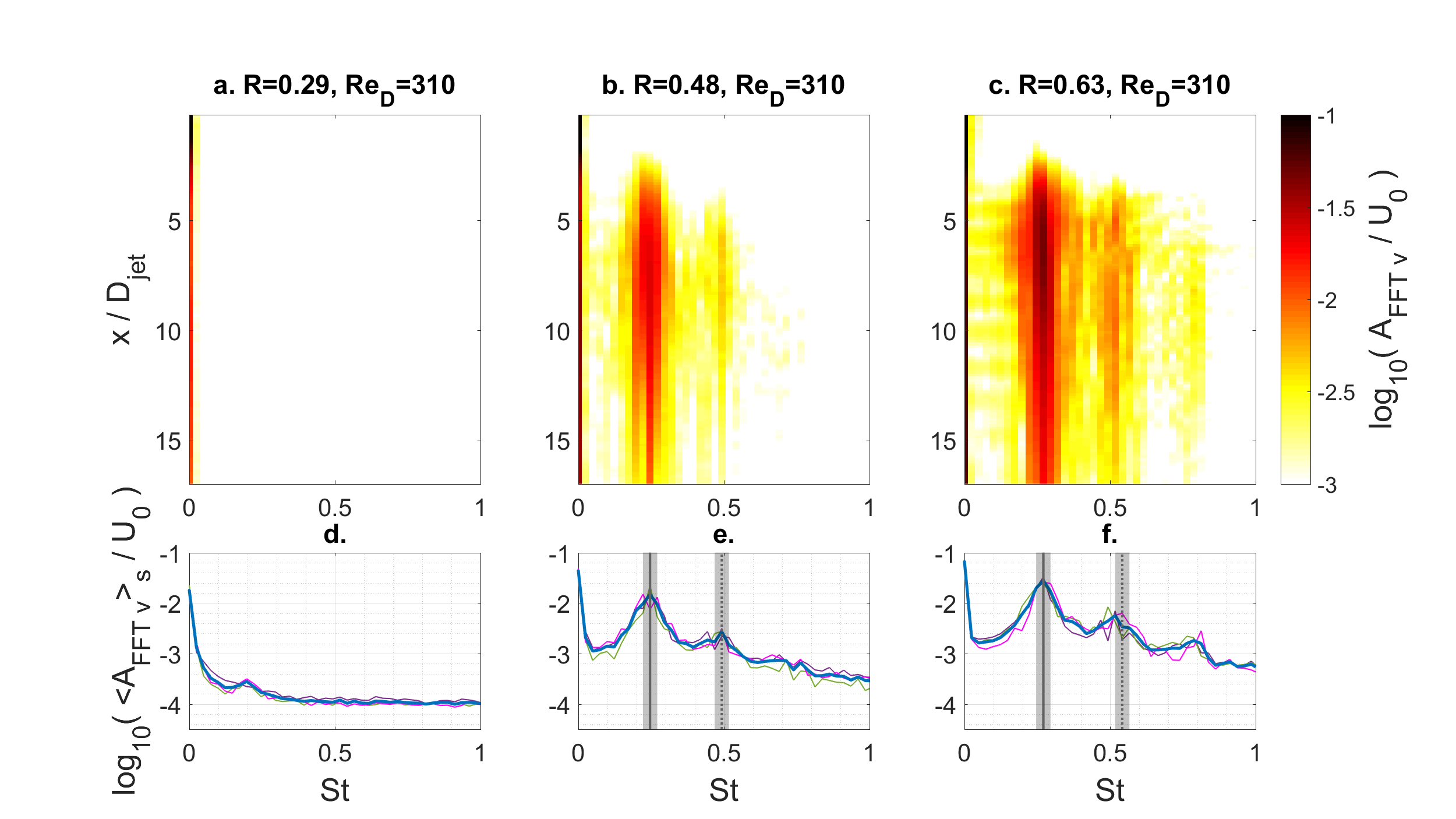}\\ 
\caption{ Ensemble-averaged FFT spectra of wall-normal velocity $v/U_0$ along the centerline trajectory for $\Rey_D=310$ and for $R=0.29$ (a), $R=0.48$ (b), and $R=0.63$ (c); the same FFT spectra averaged along $s$ (thick blue line in d,e,f) supplemented with similar spectra computed for three different realizations separately (thin dashed-dotted magenta, green and violet lines). All plots are in semi-logarithmic scale and each column corresponds to the same $R$; the fundamental frequency of hairpin shedding ($f_{HS}$) and its first harmonic ($2f_{HS}$) are indicated by solid and dashed vertical gray lines, respectively. Shaded regions indicate the interval that corresponds to the double of the spectral resolution of FFT transform.} 
\label{fig:Spectral1}   
\hspace{-0.2cm} \includegraphics[trim={0 1.65cm 0 -1.5cm},scale=0.32]{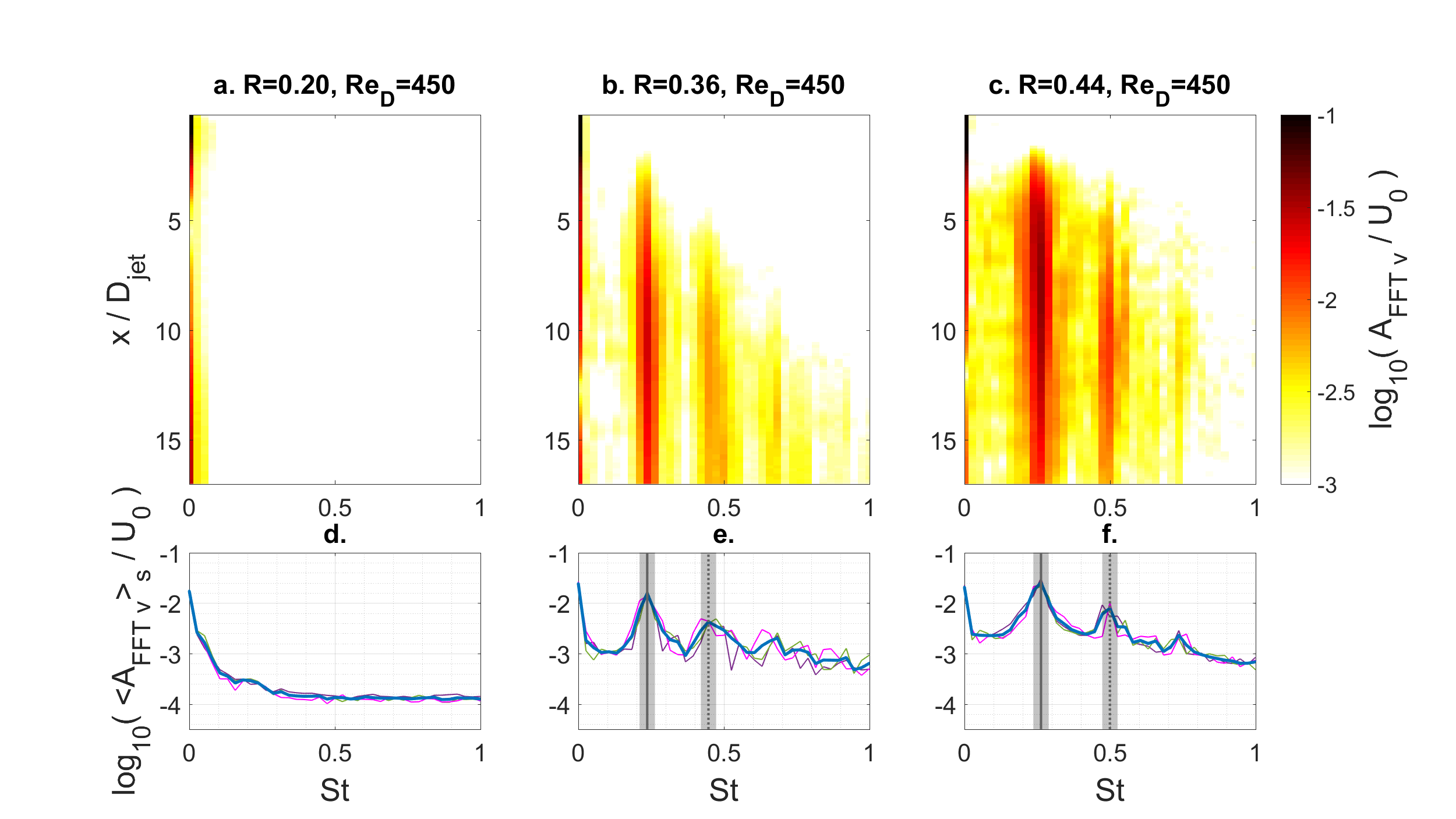}\\ 
\caption{ The same as Fig. \ref{fig:Spectral1} but for $\Rey_D=450$ and for $R=0.20$ (a), $R=0.36$ (b), and $R=0.44$ (c).} 
\label{fig:Spectral2}  
\end{figure}

\begin{figure}
\hspace{-0.2cm} \includegraphics[trim={0 1.65cm 0 -1.5cm},scale=0.32]{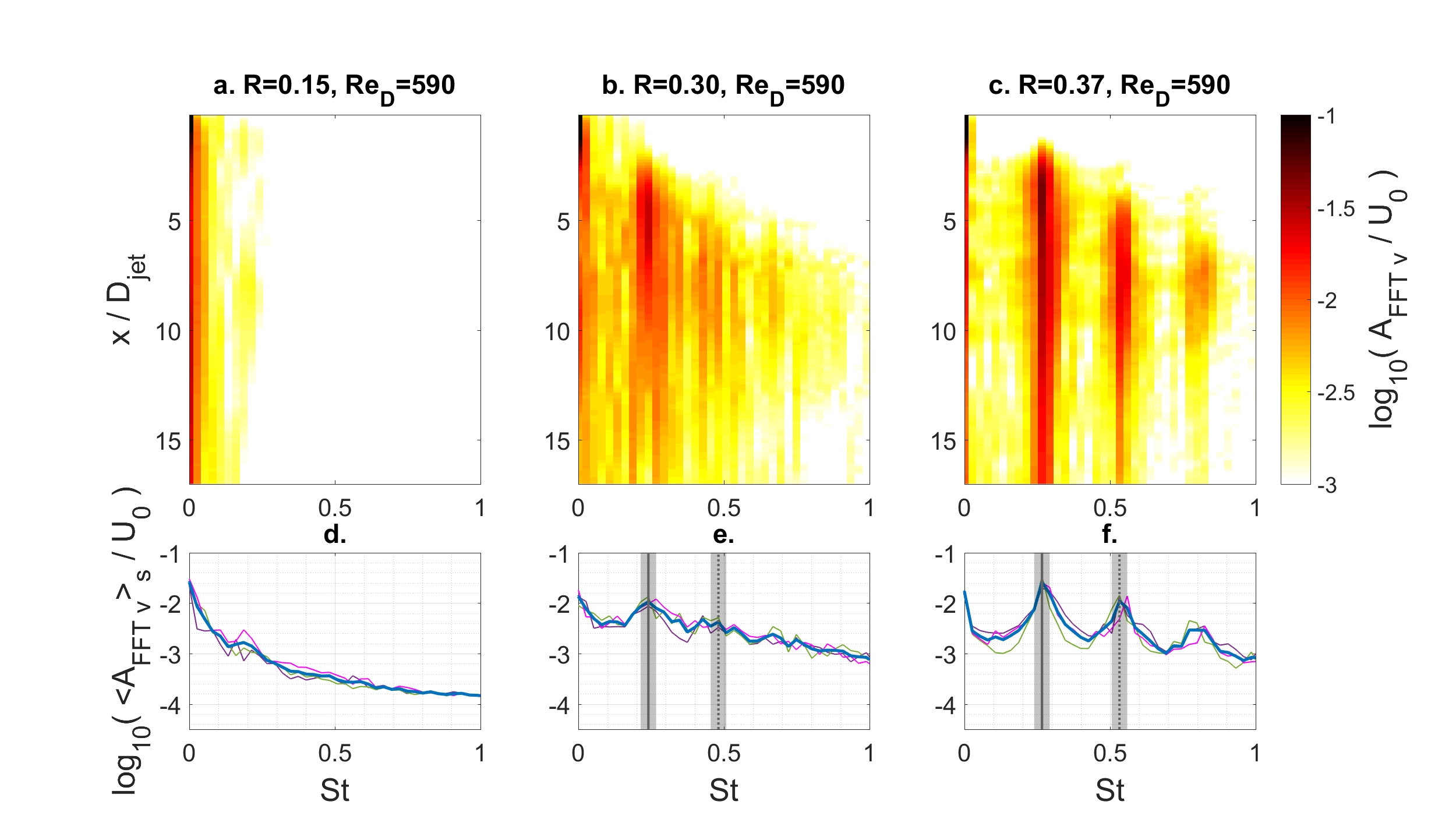}\\ 
\caption{ The same as Fig. \ref{fig:Spectral1} but for $\Rey_D=590$ and for $R=0.15$ (a), $R=0.30$ (b), and $R=0.37$ (c).} 
\label{fig:Spectral3}  
\hspace{0.55cm} \includegraphics[trim={0 1.65cm 0 -0.80cm},scale=0.32]{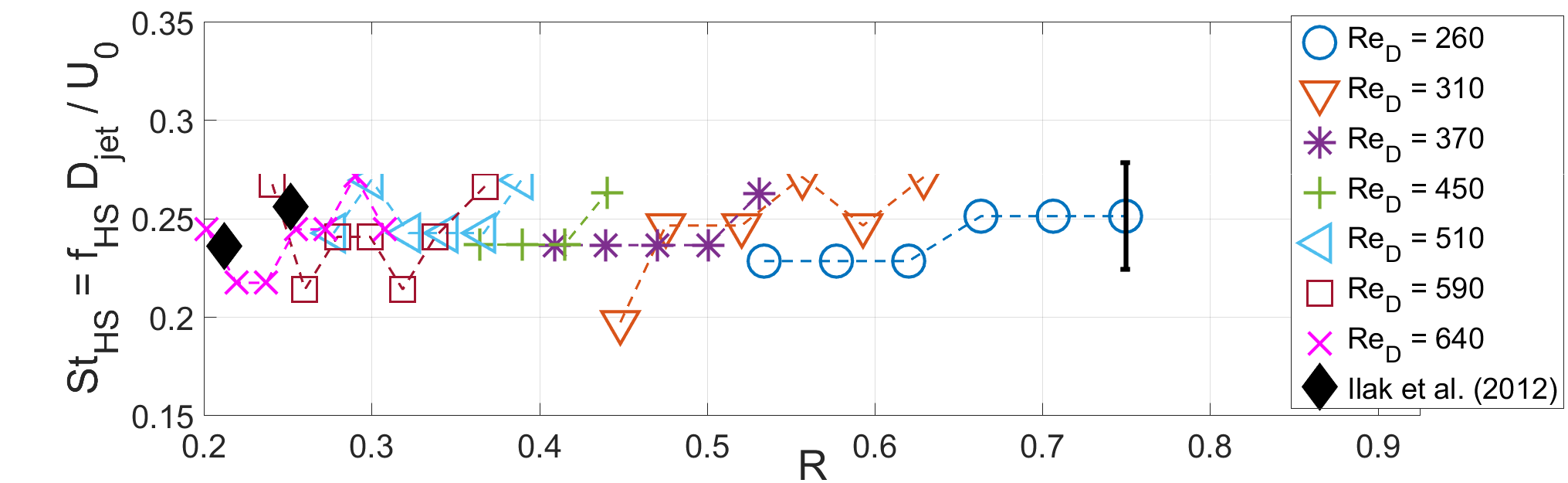}\\ 
\caption{ Dependence of $St_{HS}$ on $R$. Values of $\Rey_D$ are indicated in the legend.}

\label{fig:Freq1}  
\hspace{-1.20cm} \includegraphics[trim={0 1.65cm 0 -0.80cm},scale=0.32]{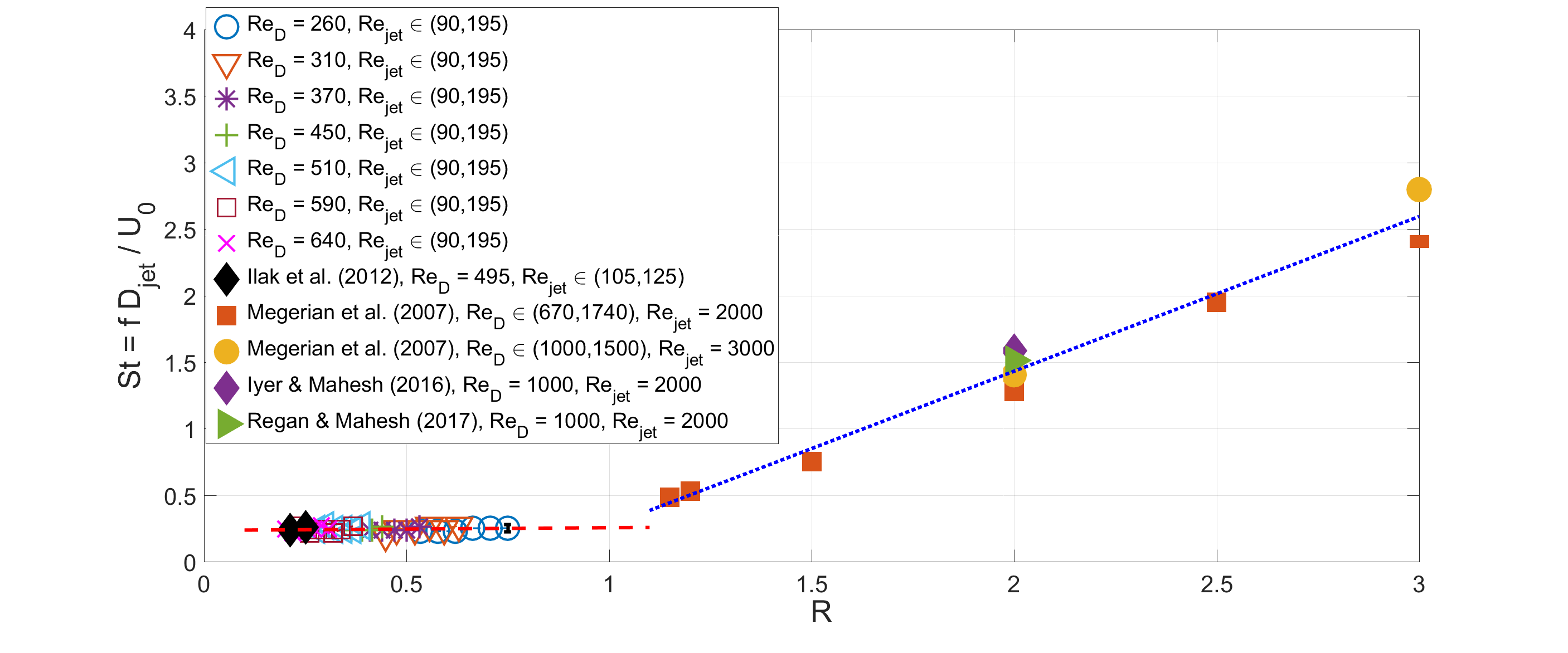}\\ 
\caption{ Comparison of the data from Fig. \ref{fig:Freq1} with the results in high-velocity-ratio regime reported by \citet{megerian_transverse_jet_2007,iyer_numerical_2016,regan_global_2017}. Range of $\Rey_D$ and $\Rey_{jet}$ for each study is indicated in the legend.} 

\label{fig:Freq2} 
\end{figure}

To characterise hairpin shedding instability, we calculate the one dimensional Fast Fourier Transform of the time evolution of the wall-normal velocity component $v/U_0$. We separately analyze each spatial location within our measurement region and each set of measurements. Then, for each combination of $R$ and $\Rey_D$ we ensemble-average the spectra over three realizations. In Fig. \ref{fig:Spectral1}-\ref{fig:Spectral3} we present the results for $\Rey_D=310,450,590$. The ordinate corresponds to the streamwise coordinate of the jet centerline trajectory and the abscissa represents the Strouhal number $St=fD_{jet}/U_0$. When $R$ is high enough, distinct single-tone oscillations with fundamental frequency and higher harmonics can be clearly distinguished (Fig. \ref{fig:Spectral1}-\ref{fig:Spectral3}b,c), which is a signature of self-sustained oscillations of hairpin shedding vortices. Hairpins are generated at some distance from the jet orifice, and once they are formed, the spectral behaviour does not change along centerline trajectory, showing a global mode structure. For comparison, in Fig. \ref{fig:Spectral1}-\ref{fig:Spectral3}a we show the spectra for the flow condition at which the distinct, self-sustained hairpin shedding was not observed.

To complete our analysis, we averaged each spectrum shown at the top of Fig. \ref{fig:Spectral1}-\ref{fig:Spectral3} over the centerline trajectory and the results are plotted as a thick blue line at the bottom of Fig. \ref{fig:Spectral1}-\ref{fig:Spectral3}. Individual realizations (shown as thin dashed-dotted magenta, green and violet lines at the bottom of Fig. \ref{fig:Spectral1}-\ref{fig:Spectral3}) and ensemble-averaged spectra overlap, as expected. At the bottom of Fig. \ref{fig:Spectral1}-\ref{fig:Spectral3} we mark a spectral peak that corresponds to hairpin shedding frequency $St_{HS} = f_{HS} D_{jet}/ U_0$ and its first harmonic $2St_{HS}$. Shaded regions correspond to the double of the resolution of FFT transform, which is about $0.027$.

With the procedure shown in Fig. \ref{fig:Spectral1}-\ref{fig:Spectral3} we determine the dependence of $St_{HS}$ on $R$ and $\Rey_D$. Fig. \ref{fig:Freq1} highlights that Strouhal number is nearly constant ($St_{HS} = 0.24\pm 0.02$, error based on standard deviation) for the investigated range of parameter space. This implies that the frequency of hairpin shedding $f_{HS}$ grows linearly with $\Rey_D$. Each colour in Fig. \ref{fig:Freq1} corresponds to different value of $\Rey_D$ and total length of error bar  represents the double of spectral resolution of FFT tranform. \citet{ilak_bifurcation_2012} reported the Strouhal number for $\Rey_D=495$ and two different velocity ratios $R^*=0.675$ and $R^*=0.8$. However, since their definition was based on the jet velocity, we need to multiply their reported values by $R^*$ to compare with our data. This results in $St_{HS} = 0.236$ for $R^*=0.675$ and $St_{HS} = 0.256$ for $R^*=0.8$, which is in close agreement with our value (black filled diamonds in Fig. \ref{fig:Freq1}). Note that in this figure we also transform their velocity ratio $R^*$ based on the jet centerline velocity to $R$ ($R=0.314R_{*}$ for their jet velocity profile).

\begin{figure}
\hspace{0.3cm} \includegraphics[trim={0 1.65cm 0 -0.64cm},scale=0.29]{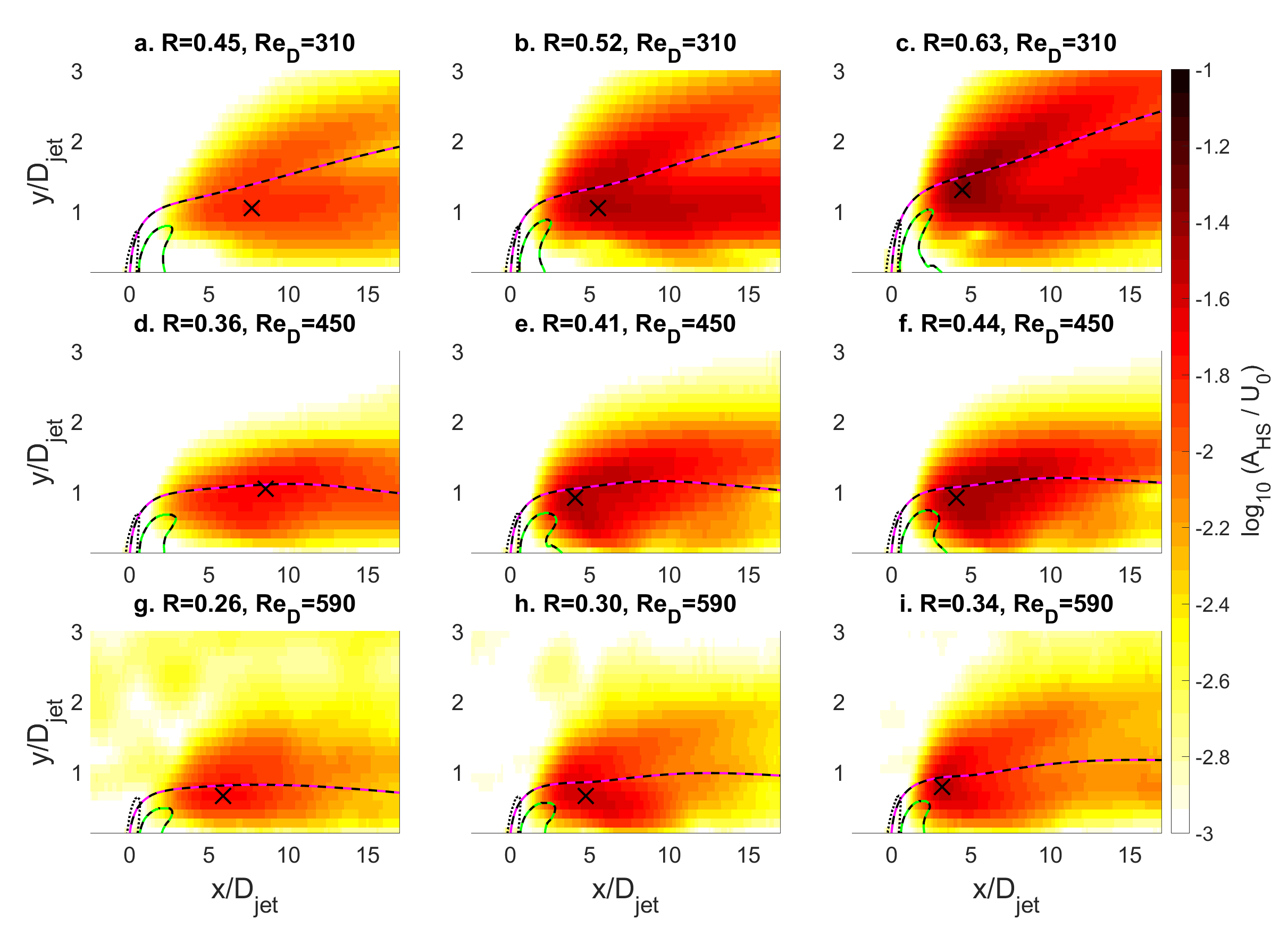}\\
\caption{ Spatial distribution of the hairpin amplitude (global mode) corresponding to the hairpin shedding frequency ($St_{HS}$), calculated from the normalized ensemble-averaged FFT spectrum for $\Rey_D = 310$ (a,b,c), $\Rey_D = 450$ (d,e,f), and $\Rey_D = 590$ (g,h,i). Jet centerlines (magenta-black dashed curves), downstream recirculation zones (green-black dashed curves) and jet contour that corresponds to the half of maximal measured jet vertical velocity (black dashed curves) are also shown. Black x symbols mark the position of the global spatial maximum of the global mode.} 
\label{fig:GlobalMode1} 

\hspace{1.35cm} \includegraphics[trim={0 1.65cm 0 -1.16cm},scale=0.29]{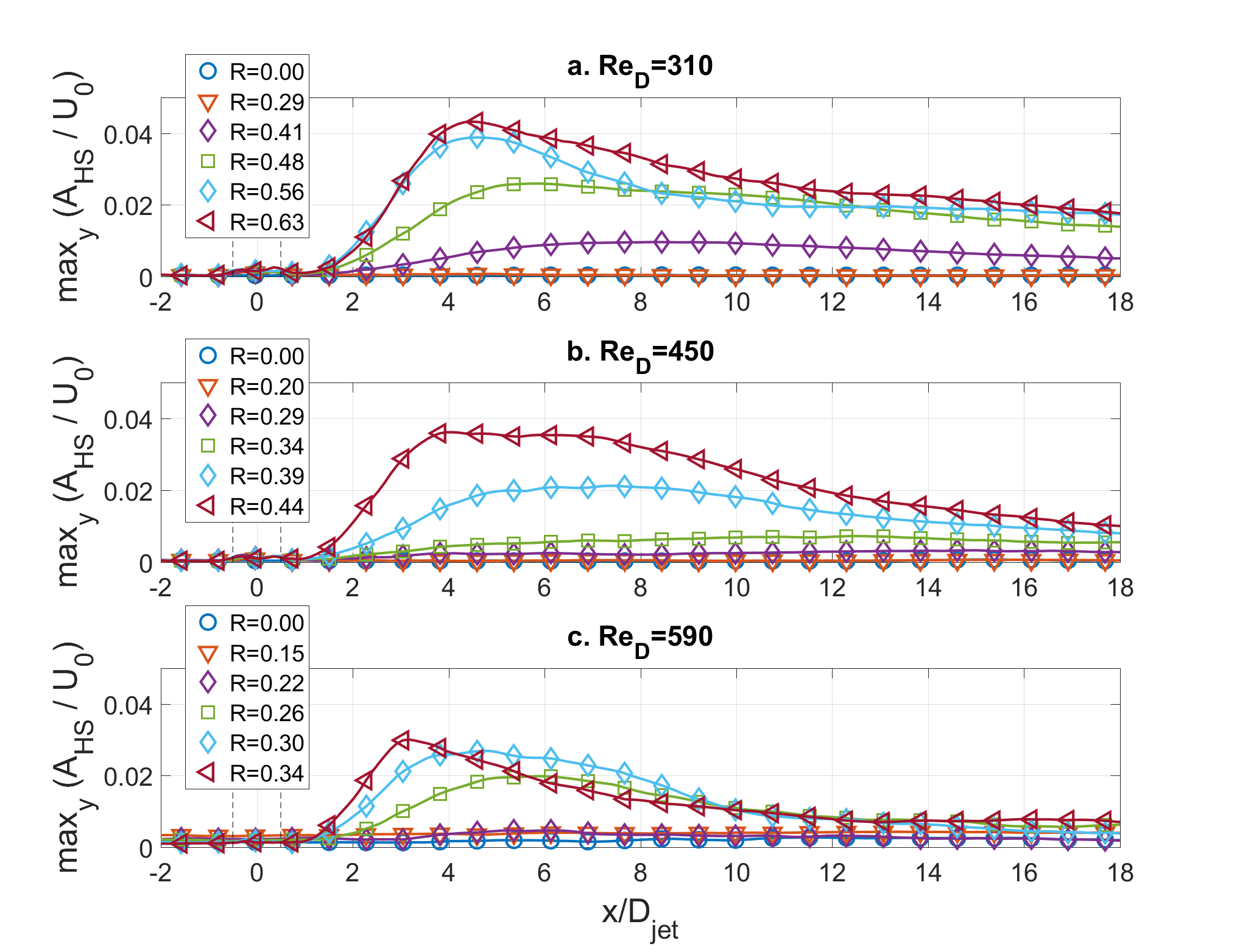}\\ 
\caption{ Streamwise evolution of the hairpin global mode for $\Rey=310$ (a), $\Rey=450$ (b), and $\Rey=590$ (c). Each point at a given $x/D_{jet}$ is obtained as maximal value of global mode of hairpin instability (Fig. \ref{fig:GlobalMode1}) along the wall-normal direction. The black vertical dashed lines indicate the upstream and downstream edge of the jet orifice ($x/D_{jet} = -0.5$ and $x/D_{jet} = 0.5$). Actual spatial resolution is six time larger than spacing of the markers on each curve.} 
\label{fig:GlobalModeStream}  
\end{figure}

\citet{megerian_transverse_jet_2007} reported similar single pure-tone oscillations in high-velocity-ratio regime. We inferred the Strouhal numbers reported for strong single-tone oscillations for flush nozzle from their Fig. 12 (open symbols) to compare with our data. We also included in the comparison the values reported by \citet{iyer_numerical_2016} and \citet{regan_global_2017}. Since the authors used the Strouhal number $St$ based on jet velocity, we multiply inferred values of $St$ by the corresponding velocity ratio $R$. The resulting values are plotted in Fig. \ref{fig:Freq2} together with the data presented in Fig. \ref{fig:Freq1}. The Strouhal number that characterizes the roll-up of the upper jet shear layer seems to follow a distinct linear trend when normalized with $U_0$. However, the increase with $R$ of the characteristic Strouhal number of the upper jet shear layer vortices is substantially larger (blue dotted line) when compared to the trend predicted from our data (red dashed line, $dSt_{HS}/dR \simeq 0.02$). Moreover, the authors studied the periodic shedding of vortices of positive spanwise vorticity in the upstream jet shear layer, which is absent in our case (see Fig. \ref{fig:StreamVelocityEvolution}). Finally, \citet{megerian_transverse_jet_2007} reported a transition scenario from absolutely unstable global self-sustained oscillations in JICF (for control parameter $R < 3.1$) to convectively unstable free-jet flow ($R \rightarrow \infty$), whereas we report the transition from the flow without self-sustained hairpin shedding ($0 < R < R_{cr}$, where $R_{cr}$ is the onset of hairpin instability) to the flow with the dynamics dominated by the hairpins in the downstream near field of JICF (for $R > R_{cr}$).

We also study the global mode describing the dynamics of hairpin vortices. For each spatial location $(x,y)$ we select from 1D FFT ensemble-averaged spectrum the spectral mode $A_{FFT\,v}/U_0$ that corresponds to the characteristic Strouhal number of hairpin shedding ($St_{HS}$, indicated by shaded region in Fig. \ref{fig:Spectral1}-\ref{fig:Spectral3}). As a result we obtain the spatial distribution of the intensity of hairpin shedding instability $A_{HS}(x,y)/U_0$. Its dependence on the control parameter $R$ is shown in logarithmic scale for $\Rey_D = 310$ (Fig. \ref{fig:GlobalMode1}a,b,c), $\Rey_D = 450$ (Fig. \ref{fig:GlobalMode1}d,e,f), and $\Rey_D = 590$ (Fig. \ref{fig:GlobalMode1}g,h,i). For higher $\Rey_D$ hairpin shedding occurs at lower $R$. On each subplot we superimpose the centerline trajectory (magenta-black dashed line), downstream recirculation zone (green-black dashed line) and jet contour that corresponds to the half of maximal measured vertical velocity of the jet (black dashed curves). The maximal amplitude of hairpins (marked by black x symbols) is located in the region bounded from the top by centerline trajectory and from the bottom/left by the downstream recirculation zone. To better illustrate the spatial evolution of $A_{HS}(x,y)/U_0$ in the streamwise direction as a function of $R$ and $\Rey_D$, we calculate the maximal value of the global mode $A_{HS}(x,y)/U_0$ along the wall-normal direction ( max$_y$($A_{HS}(x,y)/U_0$) ) for each $x$ location. The resulting curves are shown in linear scale in Fig. \ref{fig:GlobalModeStream} for $\Rey_D = 310$ (a), $\Rey_D = 450$ (b), and $\Rey_D = 590$ (c), respectively. For a given $\Rey_D$ the global spatial maximum of hairpin amplitude increases with $R$ and its location shifts closer to the jet orifice. Additionally, the spatial growth of the global mode is followed by eventual decay, similar to the global mode evolution observed in the wake behind a cylinder \citep{goujon-durand_downstream_1994,zielinska_spatial_1995,wesfreid_global_1996} or three-dimensional bluff bodies \citep[][and references therein]{ormieres_transition_1999,klotz_experimental_2014}.  

\begin{figure}
\hspace{1.15cm} \includegraphics[trim={0 1.65cm 0 -0.5cm},scale=0.32]{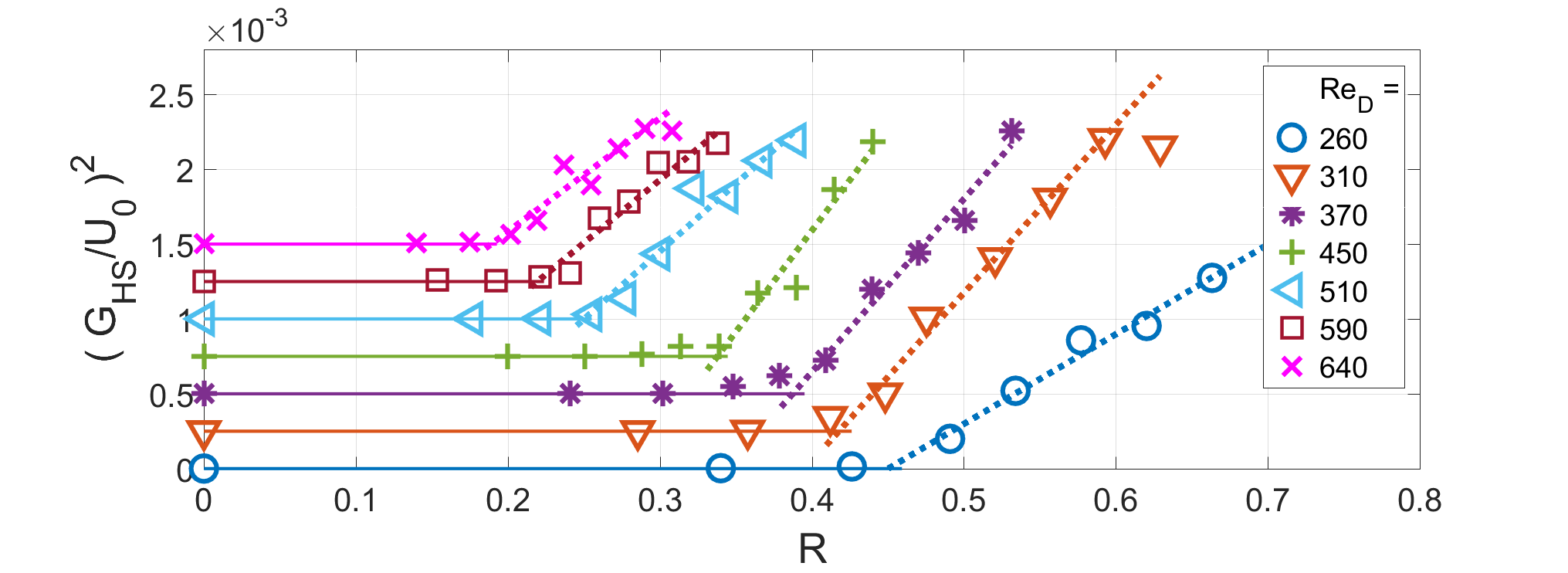}\\ 
\caption{ Bifurcation diagram of the normalized FFT amplitude of hairpin shedding $G_{HS}$ as a function of $R$ and for different $\Rey_D$. Squared amplitude is linearly dependent on control parameter $R$, as predicted by Landau equation. Each following $\Rey_D$ is shifted by $2.5\cdot10^{-4}$ upward to increase readability.}  
\label{fig:BifDiagram}  

\hspace{1.25cm} \includegraphics[trim={0 1.65cm 0 -0.8cm},scale=0.32]{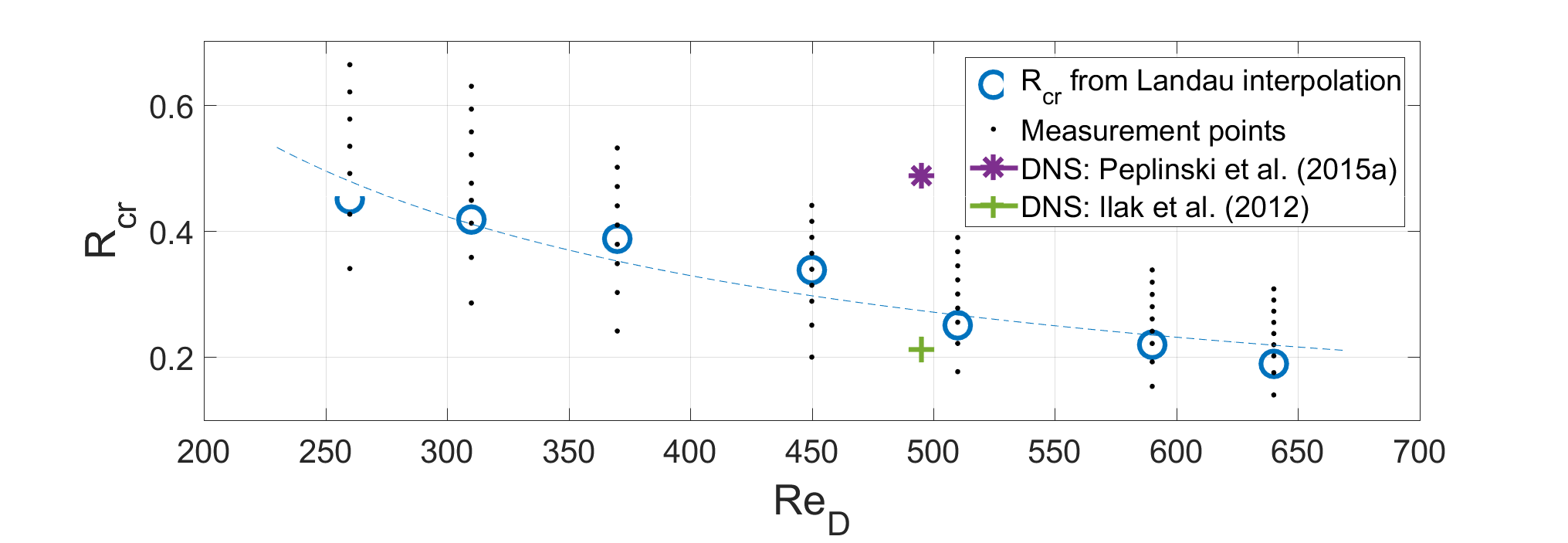}\\ 
\caption{ Dependence of critical threshold $R_{cr}$ of hairpin shedding instability as a function of $\Rey_D$ (blue circles) and measurement points (small black dots). For comparison we also present DNS results of KTH group: \citet[][purple star]{peplinski_investigations_2015} and \citet[][green cross]{ilak_bifurcation_2012}. \vspace{0.4cm}}  
\label{fig:CriticalThreshold}  
\end{figure}

\begin{figure}
\hspace{0.1cm} \includegraphics[trim={0 1.75cm 0 -0.65cm},scale=0.32]{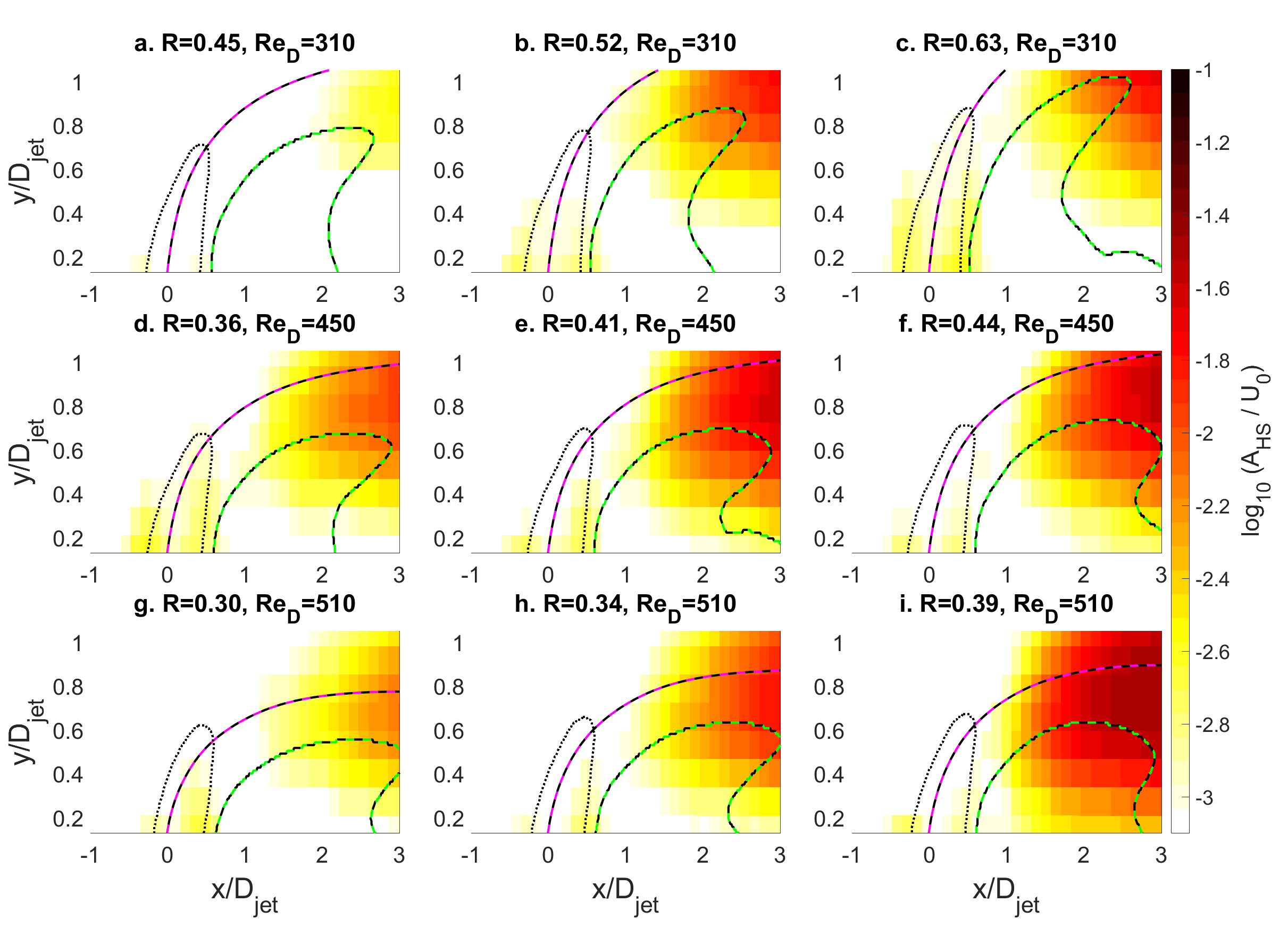}\\ 
\caption{ Similar quantity as in Fig. \ref{fig:GlobalMode1} but focused on the region close to the jet orifice: $x/D_{jet}\in (-1,3)$ and $y/D_{jet}\in (0,1.1)$.}
\label{fig:GlobalMode2}   

\hspace{1.05cm} \includegraphics[trim={0 1.65cm 0 0.11cm},scale=0.32]{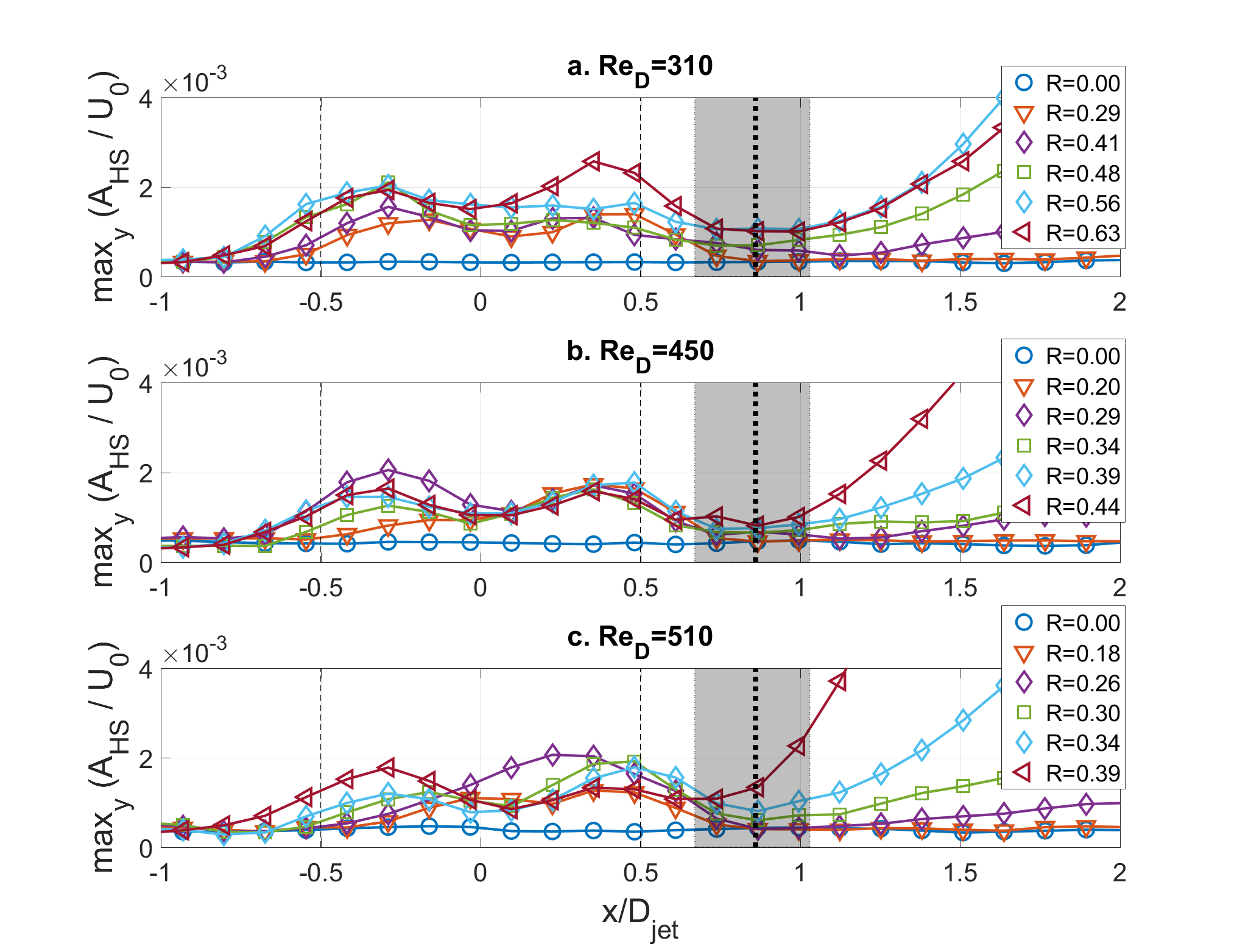}\\ 
\caption{ Similar quantity as in Fig. \ref{fig:GlobalModeStream} but focused on the region close to the jet orifice. Two weak local peaks close to $x/D_{jet}=-0.5$ and $x/D_{jet}=0.5$ correspond to the upstream and downstream jet shear layers. The black vertical dashed lines indicate the upstream and downstream edge of the jet orifice ($x/D_{jet} = -0.5$ and $x/D_{jet} = 0.5$). The black thick dotted line indicates the local minimum in streamwise direction of the global mode of hairpin instability.}  
\label{fig:GlobalModeStreamZoom}  
\end{figure}

In order to determine the critical values $R=R_{cr}$, at which the transition to hairpin shedding occurs, we apply the method described by \citet{goujon-durand_downstream_1994,zielinska_spatial_1995} in the context of {B}{\'e}nard-von {K}{\'a}rm{\'a}n street. Specifically, for each realization of combinations of ($R,\Rey_D$), we determine the global spatial maximum of the global mode of hairpin shedding instability. Then, for each ($R,\Rey_D$) pair, we calculate the mean over all realizations, to which we will refer as $G_{HS}$. We choose $R$ as the control parameter and $G_{HS}/U_0$ as an order parameter. We consider the evolution of the square of the order parameter for a fixed $\Rey_D$, which results in 7 different bifurcation diagrams (see Fig. \ref{fig:BifDiagram}). Each colour corresponds to a different $\Rey_D$ and their specific values are given in the legend. Each subsequent bifurcation diagram is shifted upwards by $2.5\cdot 10^{-4}$ to increase readability. When $R<R_{cr}$ the square of the order parameter is approximately zero, and for $R>R_{cr}$ it grows linearly with the distance from the threshold, $\sim (R-R_{cr})$. This behaviour is typical for a supercritical Hopf bifurcation and is similar for all investigated $\Rey_D$. For each $\Rey_D$ we interpolate this linear trend to zero and determine the precise value of the threshold $R_{cr}$ (see Fig. \ref{fig:CriticalThreshold}). $R_{cr}$ monotonically decreases with $\Rey_D$ and can be approximated by the power law $R_{cr}\sim\Rey_D^{-0.9}$. Similar trend was observed by KTH group \citep{chauvat_global_2017}. For comparison, we also plot on the same figure other DNS results of KTH group: \citet[][purple star]{peplinski_investigations_2015} and \citet[][green cross]{ilak_bifurcation_2012}.

We also note that a similar linear increase of the squared amplitude of the upstream jet shear layer roll-up with the distance from the instability threshold was already observed by \citet{davitian_transition_2010} in high-jet-velocity regime. The authors considered the same control parameter as in our case (jet velocity ratio $R$) and different order parameter (r.m.s. of wall-normal velocity fluctuations normalized with the jet velocity $V_{jet}$). However, they studied the evolution of the amplitude perturbations for a fixed distance from the jet orifice along the upstream jet shear layer (equivalent to method used by \citet{mathis_benard-von_1984} for the wake behind a cylinder) and obtained different values for the critical parameter depending on the spatial location. In contrast, we consider the global spatial maximum of the global mode of hairpin instability as an order parameter, similarly to \citet{goujon-durand_downstream_1994,wesfreid_global_1996,klotz_experimental_2014}.

In Fig. \ref{fig:GlobalMode2} we present the spatial distribution of the global mode of hairpin instability $A_{HS}(x,y)/U_0$ focused on the region close to the jet orifice ($x/D_{jet}\in (-1,3)$ and $y/D_{jet}\in (0,1.1)$) to distinguish between the jet shear layers and hairpin instability. Similar to Fig. \ref{fig:GlobalMode1}, magenta-black dashed, green-black dashed and black dotted curves represent the centerline trajectory, downstream recirculation zone and jet contour corresponding to the half of maximal measured vertical velocity of the jet, respectively. One can observe two regions of local weak increase of the amplitude of the hairpin global mode $A_{HS}(x,y)/U_0$ in the vicinity of the upstream ($x/D_{jet}=-0.5$) and downstream ($x/D_{jet}=0.5$) edges of the jet. Further downstream these peaks are followed by substantial increase of $A_{HS}(x,y)/U_0$ due to the formation of hairpin vortices. In Fig. \ref{fig:GlobalModeStreamZoom} we plot the quantity max$_y$($A_{HS}/U_0$) to illustrate in more details the spatial evolution of the global mode of hairpin instability. For supercritical conditions ($R>R_{cr}$) the quantity max$_y$($A_{HS}/U_0$) does not grow monotonically in the streamwise direction, i.e. two local peaks close to the upstream and downstream jet edges can be distinguished. However, their amplitude is an order of magnitude lower than the global spatial maximum of the hairpin global mode (compare with Fig. \ref{fig:GlobalModeStream}). These local peaks are followed in the downstream direction ($x/D_{jet}>0.5$) by a local minimum that we use as the criterion to define the hairpin formation. We determine its location ($x_{HS}$) for each combination of $R$ and $\Rey_D$. By ensemble-averaging over these combinations we obtain $x_{HS}/D_{jet} = 0.85 \pm 0.18$. The error is based on STD of $x_{HS}$ for all combinations and its value is of the same order as the spatial resolution of our measurements. The resulting location of $x_{HS}$ is marked by a vertical solid black line in Fig. \ref{fig:GlobalModeStreamZoom} and the shaded region corresponds to estimated error. It is also to be noted that the very weak local maxima of upstream and downstream jet shear layers are present even for subcritical case ($0<R<R_{cr}$, orange triangles in Fig. \ref{fig:GlobalModeStreamZoom}a,b,c), for which self-sustained hairpin shedding from the downstream recirculation zone was not observed.

\section{Conclusions}

We present a detailed experimental study of JICF in low-velocity-ratio regime, varying both crossflow and bulk jet speed, which spans the region of $R\in(0.14,0.75)$ and $\Rey_D\in(260,640)$ in the parameter space. Our analysis is mainly focused on the transition to hairpin shedding state and its instability properties. We demonstrate that using the free-stream velocity and jet diameter for normalization, the characteristic Strouhal number of hairpins can be approximated by $St_{HS} = 0.24 \pm0.02$ for the investigated range of parameters. Dynamical analysis reveals that the square of the hairpin velocity fluctuations grow linearly with velocity ratio $R$. This result is predicted by the Landau model describing the weakly nonlinear saturation of the state after transition, which shows quantitatively that the hairpin instability occurs through supercritical Hopf bifurcation. Using this dependence, we determine the threshold of hairpin shedding instability for the range of $\Rey_D$ studied here and show that it decreases monotonically with $\Rey_D$. In addition, the dynamics of hairpin vortices measured experimentally, as well as their characteristic Strouhal number, are in good qualitative agreement with numerical simulations of \citet{ilak_bifurcation_2012} for supercritical range of velocity ratio $R>R_{cr}$. Our results demonstrate that for the considered range of parameters, the dynamics of self-sustained hairpin shedding in JICF is an analogue to other hydrodynamical oscillators such as wake behind different bluff bodies. The structure of shed hairpins is similar to the three-dimensional bluff bodies (i.e traveling wave) and the global mode has a similar spatial evolution in the downstream direction: after initial amplification of the hairpins, their amplitude eventually decays as they are advected downstream. Additionally, global spatial maximum of the amplitude of the hairpin global mode increases and shifts closer to the jet orifice when R is increased.

\section*{Acknowledgements}
We thank Jacek Rokicki and Jos\'{e} M. Lopez for fruitful discussion, and Guillaume Saingier for help with experiments. L.K. was partially supported by the European Union’s Horizon 2020 research and innovation programme under the Marie Sk\l odowska-Curie grant agreement No 754411.
\bibliographystyle{jfm}
\bibliography{Bibliography4}

\end{document}